%% file: main.tex
\documentclass[10pt]{iopart}

\usepackage{iopams}
\expandafter\let\csname equation*\endcsname\relax
\expandafter\let\csname endequation*\endcsname\relax
\usepackage{epstopdf}
\usepackage{graphicx}
\usepackage{textcomp}
\usepackage{color}
\usepackage[version=4]{mhchem}
\usepackage{pgfplotstable}
\usepackage{array}
\usepackage{tabularx}
\usepackage{booktabs}
\usepackage{threeparttable}
\usepackage{pdflscape}
\usepackage{capt-of}

\newcolumntype{L}{>{\raggedright\arraybackslash}X}
\newcolumntype{C}[1]{>{\centering\arraybackslash}p{#1}}

\newcommand{\Qi}{Q_{\mathrm{i}}}
\pgfplotsset{compat=1.18}

\usepackage{siunitx}
\usepackage{amsmath}
\usepackage[T1]{fontenc} %resolve polish letter stuff in references
\usepackage{hyperref} %just added to make reference checking easier
\usepackage{doi}
\usepackage[numbers]{natbib}

\begin{document}
\newcommand{\unitm}[1]{\ensuremath{\,\mathrm{#1}}}
\newcommand{\LN}{LiNbO\textsubscript{3}}
\newcommand{\LT}{LiTaO\textsubscript{3}}

\topical%[WGM nonlinear and quantum optics]
{Nonlinear and Quantum Optics with Whispering Gallery Resonators Revisited}

\author{Dmitry V Strekalov\textsuperscript{1,2}, 
Nicholas J Lambert\textsuperscript{1,2}, 
Christoph Marquardt\textsuperscript{3,4}, Andrey B Matsko\textsuperscript{5}, 
Harald G L Schwefel\textsuperscript{1,2}, 
Florian Sedlmeir\textsuperscript{1,2,6}, 
Mallika I Suresh\textsuperscript{1,2}, and 
Gerd Leuchs\textsuperscript{3,4}}

\address{\textsuperscript{1}Department of Physics, University of Otago, Dunedin, New Zealand}
\address{ \textsuperscript{2}Dodd-Walls Centre for Photonic and Quantum Technologies, New Zealand}
\address{ \textsuperscript{3}Max Planck Institute for the Science of Light, G\"{u}nther-Scharowsky-Stra\ss e 1/Building 24, 91058 Erlangen, Germany}
\address{ \textsuperscript{4}Institute for Optics, Information and Photonics, University Erlangen-N\"{u}rnberg, Staudtstr.7/B2, 91058 Erlangen, Germany}
\address{\textsuperscript{5}Jet Propulsion Laboratory, California Institute of Technology, Pasadena, CA 91108, USA}
\address{ \textsuperscript{6}Universidad Carlos III de Madrid,
Departamento de Teor\'ia de la Se\~nal y Comunicaciones,
Avenida de la Universidad 30,
28911 Legan\'es, Madrid, Spain}
\ead{strekalov@yahoo.com}
\vspace{10pt}
\begin{indented}
\item[]\today
\end{indented}

\begin{abstract}
In 2016 we published a review paper ``Nonlinear and Quantum Optics with Whispering Gallery Resonators"  \cite{Strekalov16rev}. While writing this review, we found it often necessary to defer addressing many challenges of the field to the future. At the same time, many potential benefits offered by nonlinear optical resonators were also deferred to the future pending the underlying technology development. The past ten years have seen a great deal of such development, including in the field of resonator optics. We feel that an update to our 2016 review is in order. %In this new review we focus on the most significant developments in the field that took place in past ten years.
 We start by acknowledging the progress in the broader area of quantum-optical technology and the increased role of micro-resonators in this area. While the stand-alone resonators remain highly relevant, the emphasis clearly shifts towards chip-integrated resonators and resonator arrays. Both approaches advance by the exploration of new optical materials and fabrication techniques, which is one important topic of our review. Then we review the progress in traditional nonlinear-optics resonator research: second- and third-order frequency conversion, Raman and Brillouin scattering, optomechanics, and lasing. Finally, we discuss the implications of these developments for quantum optics and photonics.
\end{abstract}


\vspace{2pc}
\maketitle

\tableofcontents

\ioptwocol

\section{Introduction}\label{sec:Intro}

\subsection{Nonlinear and quantum optics over past 10 years}\label{sec:Intro_nonlin}

The nonlinear and quantum optics has undergone a substantial evolution since 2016. The overall vector of this evolution can be described as mainly technology-oriented. This includes the development of new nonlinear-optical materials \cite{yan2024crystals}, improvement of the fabrication techniques, and most importantly, a shift from stand-alone optical bench devices to integrated chip-based ones.  Two characteristic examples are the thin-film silicon nitride and lithium niobate (TFLN, \cite{zhu2021LNOfilms,vazimali2022TFLN,hu2025TFLM}) photonic integrated circuits (PIC) platforms, reviewed in Section~\ref{sec:chip}. They both strongly rely on the high-$Q$ microresonators. The TFLN technology has opened up the pathway to the chip-scale optical frequency synthesizers and precision metrology systems. In the quantum optics field, TFLN enabled miniature turn-key sources of the squeezed light, entangled photon pairs, and heralded photons, as well as more advanced quantum-state manipulation devices and quantum information protocols \cite{finco2024TFLNquant,kim2025TFLNquant,karami2026quant}. This development leads to the paradigm shift from fixed to reconfigurable quantum optics instrumentation.  Particularly interesting in this context are narrow-band squeezed or entangled optical states generated in nonlinear WGM resonators. Such states can efficiently interface non-optical qubits, while allowing for the entanglement swapping between independent sources via Hong-Ou-Mandel interference \cite{huang2026HOM}. This will be discussed in Section~\ref{sec:quant}.

Optical sensors provide another example of successful photonic technology maturation leveraging microresonators \cite{jiang2020WGMRsensors,yu2021sensors,liu2021sensing}. The resonator-based optical sensors have been beneficial for accurately measuring temperature \cite{ paz-buclatin2022Tsens,zhang2024sens}, pressure, strain, electric and magnetic fields \cite{ yacoby2022sensor,yu2024magsens}, acceleration, rotation, and a chemical composition of gases and liquids \cite{zhang2018sensor,yu2022sensor,mohmad2025sensor}. The performance metrics of these various sensors is too diverse to summarize it here, and besides it is well covered in the review papers we cite. Let us however point out that perhaps the most prominent application in this field is the chemical and biological sensing aimed at the detection of single viruses and bio-molecules in the evanescent field of a high-$Q$ resonator \cite{ su2017sensor,loyez2023biosens,hao2024biosens,demory2024sensor,ming2025sensor}. 
Typically based on the resonance shift and broadening in response to the evanescent field \cite{silva2021sensor} or the material refractivity perturbation, modern photonic sensors can also leverage the optomechanical and optoplasmonic phenomena \cite{bozzola2017sensing}, as well as the mode splitting, in particular near the exceptional points \cite{yu2021sensors}. Another advanced bio-sensing technique is based on the microresonator lasing \cite{toropov2021sensor}. In this case, the laser frequency shift is measured instead of the cold-cavity line shift, which affords a greater frequency resolution and optical signal-to-noise ratio.

An impressive progress has been made in the quantum transduction. In this process, high-$Q$ resonators with strong electro-optic interaction allow for the quantum states transfer between the microwave and optical photons. Quantum transduction also can be achieved in the field of quantum optomechanics. Here, the main challenge was concerned with suppressing the thermal population of mechanical modes, which requires a deep cryogenic environment as well as a thorough decoupling of the target mode from the thermal bath. The progress in optomechanics is reflected in a number of review papers, e.g. \cite{li2021rev,barzanjeh2022optomech,wen2023cooling,croquette2023rev}. 

The higher-order cubic (Kerr) nonlinearity research has been mainly centered on the optical combs generation in microresonators, pursuing the octave-spanning phase-locked comb, as well as the narrower ultra-stable combs. Here, one thrust area has been the dispersion characterization \cite{zhang2024spectroscopy} and engineering, including by coupling several resonators. This has led to an interesting research of the soliton dynamics, particularly based on the interplay of $\chi^{(2)}$ and $\chi^{(3)}$ processes.

The spectacular breakthrough in the reduction of the nonlinear-optics and quantum instruments footprint and power consumption, together with the progress in their reliability, enabled a series of airborne or satellite applications and technology demonstrations. 

\subsection{Quantum optics goes to space}

An unmistakable trend of the past decade has been the expansion of quantum optics into space \cite{bedington2017QKDsat}. The Chinese quantum space program is one of the leaders in the field. To name just a few of its efforts, Micius satellite \cite{lu2022micius} was launched on August 16, 2016 with the mission to demonstrate the satellite to ground quantum key distribution (QKD), uplink quantum teleportation, and downlink entanglement distribution. Space-to-ground QKD with a kilohertz key rate over a 1200 km distance was soon reported \cite{Liao17Micius}.  A successful satellite-mediated entanglement distribution was reported \cite{Yin17spaceBell} between two ground locations 1200 km apart, enabling a reliable violation of Bell's inequalities. Sharing the entanglement enabled a quantum state transfer across the same distance. \cite{li2022Teleport}. Space-to-ground QKD was reported also with other satellites, e.g. \cite{takenaka17QKD,Liao17QKD}.

The European quantum space program encompasses the quantum communications \cite{Oi17CubeSat,Khan18OPN}, quantum sensing, and the fundamental quantum-optics study in space \cite{restuccia2019Photon}. ESA's EAGLE-1 \cite{calistro-rivera2024EAGLE-1,hiemstra2025EAGLE-1} and SAGA missions are geared to be Europe's first operational space-based QKD systems. EAGLE-1 is furthermore viewed as a precursor to the broader European Quantum Communication Infrastructure, which will integrate fiber- and satellite-based quantum networks. The quantum sensing thrust is best represented by Bose-Einstein Condensate and Cold Atom Laboratory project \cite{frye2021BECAL}, jointly developed by Germany's DLR and USA's NASA.

A success of these rapidly developing quantum space programs inspired a publication of a multi-institutional ``manifestos" for quantum optics in space \cite{kaltenbaek2021quant_space,mohageg2022DSQL}, summarizing potential space missions aimed to use quantum optics for quantum communication, sensing, and exploration of various open questions of cosmology.

At the heart of the most ambitious of these programs is a quantum light source. This source needs to be light enough to be compatible with the space flights; sturdy enough to survive a rough launch \cite{tang16post-expplosion}; and efficient enough to use a low-power pump. This is where optically nonlinear micro-resonators come into play.

\section{Stand-alone resonators}\label{sec:fab}

Individually fabricated stand-alone crystalline WGM resonators have been historically renown for their exceptional optical quality factors, surpassing those of most integrated structures. This is why in spite of a rapidly growing interest in the chip-integrated optical resonators (discussed in the next Section), the stand-alone resonators remain important and continue to develop. One outcome of this development is the ``semi-integrated" approach, when the on-chip integrated couplers and waveguides are used in a combination with separately fabricated high-$Q$ WGM resonators, see e.g. \cite{savchenkov2017PolymerCplr,liu2018waveguide}. A notable progress has been also achieved in discovering new optical materials and developing new fabrication methods   and coupling designs for the stand-alone resonators. Let us discuss these topics in more detail.

\subsection{New optical materials for different spectral domains}%\label{sec:standalone}

The identification of suitable material platforms depends on key material properties such as spectral transparency (linear absorption loss), dispersion, and the magnitude of relevant nonlinear coefficients, %e.g., $\chi^{(2)}$, $\chi^{(3)}$, electro-optic coefficients, Kerr nonlinear index (nonlinear gain), 
which together determine the efficiency and noise performance of frequency conversion processes in nonlinear optical and quantum devices. 

Fluorides such as calcium fluoride (CaF\textsubscript{2}), magnesium fluoride (MgF\textsubscript{2}), lithium fluoride (LiF), strontium fluoride (SrF\textsubscript{2}) and barium fluoride (BaF\textsubscript{2}) are among the most studied crystalline materials for mm-sized monolithic WGM resonators due to ultra-low optical losses and high Q factors making them an ideal platform for optimising fabrication techniques~\cite{lin_dependence_2018} and coupling schemes~\cite{christensen2022Distance, christensen2025rings, lin_angled_2025} in the telecommunications band. Recently there has been investigation of barium magnesium fluoride (BaMgF\textsubscript{4}, abbreviated as BMF) as a resonator material due to its broad transparency range extending into the deep ultra violet (UV)~\cite{mrokon_electrically_2025}. Breunig's group had also been the first to assess lithium tetraborate (LB4) as a WGM resonator material for the generation of the UV light at \SI{245}{nm} more than a decade ago in 2015~\cite{furst_second-harmonic_2015}. Interest in this crystal has picked up more recently as an LB4 resonator was shown with an order of magnitude better quality factor up to $2\times 10^9$ at \SI{517}{nm}~\cite{tian_cascaded_2025}. Cascaded Raman lasing was demonstrated in the same publication and broadband second harmonic generation in an x-cut LB4 disk~\cite{tian2025SH_LB4}. 

On the long-wavelength end of the optical domain, evaluation of familiar materials for use at \SI{2}{\micro\meter} has been conducted~\cite{velazquez-ibarra_looking_2025}. Novel materials have also been recently explored including heavy-metal fluorides like ZBLAN~\cite{tebeneva_ultrahigh-q_2022}, silver gallium selenide (AgGaSe\textsubscript{2}) cadmium silicon phosphide (CdSiP\textsubscript{2})~\cite{jia2018midIR, amiune_pump_2022, amiune2023Midinfrared}, crystalline germanium~~\cite{tebeneva_crystalline_2024} and chalcogenide AMTIR-1~\cite{yang_record-high-q_2025}. Crystalline silicon has also been identified recently as having potential as a material for WGM resonators at \SI{8.6}{\micro\meter}~\cite{shitikov2026Si}.

Exploration of materials with wide transparency windows in the last decade has pushed the field of mm-sized WGM resonators into the mm-wave and THz domain~\cite{prabhu_whispering_2020} as shown with silica bubble resonators~\cite{vogt_terahertz_2017} and sapphire spheres~\cite{mathai_sensing_2018} and disks~\cite{barannik_contactless_2018, gugliandolo_whispering_2021}. With inspiration from the photonics side, there has been a number of experimental demonstrations of monolithic WGM resonators at terahertz frequencies in semiconductor materials such as high-resistivity silicon spheres~\cite{vogt_fano_2017}, GaAs disks~\cite{suresh_gallium_2023}, and silicon disks~\cite{petersen_visualization_2025} including ultra-thin disks that have ultra-high Q factors exceeding \SI{1e4}{}~\cite{vogt_ultra-high_2018, vogt_subwavelength_2020}. Research in this spectral range has flourished in the last decade with optimisation of coupling~\cite{vogt_prism_2018, yuan_mode_2019, vogt_free-space_2019}, characterisation~\cite{vogt_coherent_2019, petersen_visualization_2025} and tuning~\cite{vogt_thermal_2018, yuan_tunable_2019, vogt_anomalous_2019} (similar to work done in the optical domain~\cite{foreman_dielectric_2016, trainor2018selective,azeem_dielectric_2021}), with applications in sensing~\cite{vogt_terahertz_2020, gandhi_microresonator_2022, hill_ultra-thin_2025}.

\subsection{New designs and fabrication methods for stand-alone resonators}

\begin{figure}[htbp]
\centering\includegraphics[width=8.5cm]{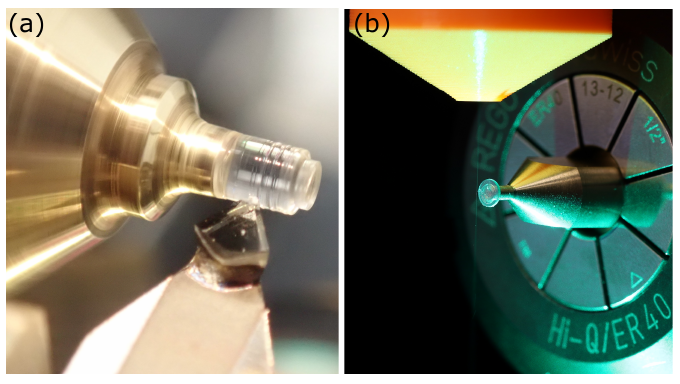}
\caption{(a) Multiple coaxial millimeter-size resonators cut using single-point diamond turning. (b) Laser ablation using pulses carrying \SI{40}{\micro J} of energy in a pulse width of $\sim$\SI{300}{fs} and central wvaelength of \SI{515}{nm}.}
\label{fig:freestanding}
\end{figure}

While scalability requirements continue to determine the direction of development of the field, individually fabricated resonators are simpler to realise at ultra-high-Q~\cite{lin_nonlinear_2017}. As such, a lot of work has been conducted in exploring and optimising their optical coupling techniques~\cite{sedlmeir_polarization-selective_2017, trainor2018selective, shitikov2018Si, tian2025SH_LB4,rosenberger2025coupler} and tuning schemes~\cite{kang_packaging_2024,yacoby_fabrication_2023}. Stand-alone resonators also occupy an important niche in applications requiring bespoke instrumentation, for example in space-deployable systems~\cite{strekalov2022Receiver,abdalmalak2022}. In such contexts, the still-unmatched performance of diamond-turned crystalline WGM resonators, particularly in terms of optical quality factor and long-term stability, offers a decisive advantage. Hence, there is still work being done into improving and/or partially automating the different fabrication steps like polishing~\cite{minkov_chemicalmechanical_2022} and annealing~\cite{yang_q-factor_2022}. Stand-alone resonators also allow more adventurous designs such as ``sandwiched'' structures for thermal compensation and regulation~\cite{lim2017sandwich,savchenkov2018sandwich} or multiple coaxial crystalline resonators~\cite{yang_kerr_2022} - similar example shown in Fig.~\ref{fig:freestanding}(a). The authors in \cite{lim2017sandwich} were able to reach the thermorefractive noise limit allowing a frequency instability of \num{1.67e-13}{} at the \SI{191}{THz} carrier, while in \cite{savchenkov2018sandwich} thermal isolation was improved by an order of magnitude with the multilayered isolation design. In~\cite{mrokon_electrically_2025}, the authors discuss the advantages of the pivoting away from mechanical diamond-cutting to ablation using femtosecond laser pulses for shaping crystalline resonators (image shown in Fig.~\ref{fig:freestanding}(b)) such as lower risks of causing cracks along cleavage planes of the crystal and significantly shorter polishing times.

Nevertheless, recent improvements in diamond turning of MgF\textsubscript{2} and CaF\textsubscript{2} resonators have yielded optical-quality surfaces without post-processing by hand-polishing, while maintaining Q factors of \SI{1e8}{}~\cite{fujii2020fab}. These advances have allowed the exploration of novel dispersion-engineered designs of crystalline resonators (with spheroidal, triangular, rectangular, and trapezoidal cross sections) for widely tunable microcombs~\cite{yang_systematic_2025}. Progress in the WGM spectrum engineering by leveraging the resonator morphology was reported by~\cite{lai2025etalon}. Using the single-mode ``photonic belt'' approach~\cite{Savchenkov06SingleMode}, the authors improved the higher-order-mode suppression by redesigning the belt geometry. As a result, a single-mode spectral etalon functionality was demonstrated in the range of 1160 to 1800 nm, enabling the ``Astrocomb''~\cite{suh2019exoplnets} spectroscopy calibration tool.

Beyond mechanically shaping the geometry, coatings on resonators have been utilised for dispersion control as well~\cite{jin_dispersion_2017} to optimise Kerr combs~\cite{inga_alumina_2020} and explore the landscape of third harmonic generation in a microbubble resonator~\cite{liu_monitoring_2025}. Metal coatings on dielectric resonators have also been of interest for applications such as electro-optic tuning~\cite{szabados_electro-optic_2020, minet_electro-optic_2021, szabados2020_SHcomb, minet2022EOlaser} and calligraphic electric field poling. This technique has seen improvements since the first demonstration in~\cite{mohageg_calligraphic_2005} by implementing a constant current control~\cite{werner_large_2017, herr2018QPM}, and was expanded to a broader list of optical crystals, such as BMF \cite{herr_fanout_2023}. 

\textit{In situ} control of a resonator radius during its fabrication was demonstrated by \cite{gil-santos2017scalable} who used cavity-enhanced photoelectrochemical etching. The resonator was coupled to a pump laser light which stimulated the etching by inducing a strong optical field at the surface. The decrease of the resonator radius detuned it from the laser and stopped the etching, allowing for a very precise eigenfrequency control.

Doping of crystalline materials for WGM resonators has attracted growing interest, with studies spanning both established laser host materials and emerging platforms for quantum technologies. These studies and their results are summarized in Table~\ref{tab:doped_crystalline_WGM}.% Early work done more than a decade ago explored praseodymium-doped yttrium orthosilicate (Pr\textsuperscript{3+}:YSO) for accessing the strong coupling regime of cavity QED~\cite{mcauslan_coherent_2011}. Fabrication and optical-loss characterisation have renewed emphasis on the suitability of YSO as a host material for rare-earth-ion-based quantum systems~\cite{norman_measuring_2022}. Building on this foundation, erbium-doped YSO (Er:YSO) WGM resonators have been employed to demonstrate optical detection of magnetic fields, which could be used as a tool for characterising atom-based microwave-to-optical transducers~\cite{ma_optically_2023}. Attention also turned to titanium-sapphire (Ti:Sapph), a material long used in bulk lasers. Lasing in a WGM resonator made from this crystal was demonstrated by~\cite{azeem2022lasing}. Lasing has also been demonstrated in a ytterbium-doped calcium fluoride (Yb:CaF\textsubscript{2}) resonator~\cite{fan_whispering_2025}.

%%%%%%%%%%%%%%%%%%%%%%%%%%
\begin{table*}[htb]
\centering
\caption{Comparison of doped crystalline whispering-gallery-mode (WGM) resonators reported in the literature.}
\label{tab:doped_crystalline_WGM}
\begin{tabular}{p{1.5cm}p{1.9cm}p{1.5cm}p{2.8cm}p{4.4cm}p{2.5cm}}
\hline\\
\textbf{Ref.} & 
\textbf{Material} & 
\textbf{Dopant} &
\textbf{Demonstrated functionality} &
\textbf{Key results} &
\textbf{Application relevance} \\
\hline\\

Minet \textit{et~al.}~\cite{minet2022EOlaser}&
LiNbO$_3$ &
Nd, MgO &
WGM-pumped single-frequency CW lasing at 1064 nm. &
Belt resonator structure reduces the WGM volume and the modes density. Electrically tunable over 3.5 GHz. &
Suitable for full integration with existing lithium niobate photonic platforms. \\

\hline\\

McAuslan \textit{et~al.}~\cite{mcauslan_coherent_2011} &
Y$_2$SiO$_5$ &
Pr$^{3+}$ &
Coherent spectroscopy and cavity-QED investigation using a rare-earth-ion-doped WGM resonator. &
Measured mode volume of $6.99\times10^{-12}$ m$^3$ and Pr$^{3+}$ coherence time of $68~\mu$s. Observation of optical bistability and normal-mode splitting associated with ion--cavity coupling. &
Strong coupling cavity QED and quantum memories based on rare-earth ions. \\

\hline\\

Herr \textit{et~al.}~\cite{herr2019tunable} &
YVO$_4$ &
Nd &
Flood-pumped CW lasing at 1064 nm.  &
Single-frequency, mode-hop-free frequency tuning over 11 GHz. 30 MHz stability. &
Low-cost, compact spectroscopy instruments.\\

\hline\\

Norman \textit{et~al.}~\cite{norman_measuring_2022} &
Y$_2$SiO$_5$ &
(RE-doped platform) &
Fabrication and optical-loss characterization of crystalline YSO WGM resonators. &
Demonstrated intrinsic quality factors up to $Q\approx1.1\times10^{9}$ at telecom wavelengths and analyzed scattering-related loss mechanisms. &
High-Q cavities for rare-earth-ion quantum technologies and coherent optical interfaces. \\

\hline\\

Ma \textit{et~al.}~\cite{ma_optically_2023} &
Y$_2$SiO$_5$ &
Er$^{3+}$ &
Optically detected magnetic resonance using an Er-doped WGM resonator. &
Demonstrated microwave-to-optical spin characterization in a cryogenic Er:YSO WGM cavity with optical quality factors exceeding $10^{8}$ and ensemble strong coupling. &
Characterization of atom-like microwave-to-optical transducers and hybrid quantum systems. \\

\hline\\

Azeem \textit{et~al.}~\cite{azeem2022lasing} &
Al$_2$O$_3$ (Ti:sapphire) &
Ti$^{3+}$ &
Whispering-gallery-mode laser operation in a crystalline gain resonator. &
Demonstrated the first Ti:sapphire WGM laser. Achieved $Q\sim10^{8}$, lasing threshold of approximately $14.2$ mW, and slope efficiency of about 34\%. &
Compact tunable solid-state lasers, nonlinear optics, and integrated photonics. \\

\hline\\

Fan \textit{et~al.}~\cite{fan_whispering_2025} &
CaF$_2$ &
Yb$^{3+}$ &
Fiber-coupled WGM microlaser based on a rare-earth-doped fluoride crystal. &
Fabricated a sub-millimeter Yb:CaF$_2$ WGM microcavity. Demonstrated 1064 nm lasing under 976 nm pumping with milliwatt-level output and low threshold operation. &
Compact infrared microlasers and integrated photonic sources. \\

\hline
\end{tabular}
\end{table*}
%%%%%%%%%%%%%%%%%%%%%%%%%

Magneto-optical tuning of WGM resonator frequencies can be used both in the context of WGM spectral design and magnetic field sensing. This phenomenon was studied by \cite{vincent2020magnetic,chai_nonreciprocity_2020,danilin2021FaradayWGM}. Particularly, \cite{chai_nonreciprocity_2020} predicted and observed a linear Faraday effect arising in both quasi-TM and quasi-TE modes due to the ``minor" electric field $E_\phi$ and leading to a magnetically sensitive splitting between the clockwise (CW) and counter-clockwise (CCW) propagating WGMs. Let us point out that the CW-CCW degeneracy can be removed not only by a magnetic field, but also by a mechanical rotation of the WGM resonator. This phenomenon enables WGM-based optical gyroscopes \cite{savchenkov2018gyro,silver2021gyro} and the non-reciprocal photonic blockade, which will be discussed below.

%Continuous study of photorefractivity in lithium niobate resonators: ``Mitigating photorefractive effect in thin-film lithium niobate microring resonators" \cite{xu2021photorefraction}. Specifically addresses resonators with dielectric cladding, pointing out that the cladding makes this effect worse than in bare resonators - not very important.

\newpage

\section{Chip-integrated resonators}\label{sec:chip}
Since roughly 2015, WGM and ring resonators have undergone a decisive transition from their stand-alone implementations toward semi and fully integrated devices on chip. Integration has become increasingly attractive due to scalability, mechanical robustness, compact footprints, and compatibility with complex photonic and microwave circuits. Beyond large scale manufacturability, integrated platforms enable strong confinement of the electric fields and allow for aggressive dispersion engineering, which are essential for enhancing and tailoring nonlinear interactions.

On chip, different resonator geometries have emerged that reflect a trade-off between optical confinement, scattering losses, and fabrication scalability. An early approach continues the philosophy of ultra-high-Q bulk resonators by transferring it to the chip: undercut microdisks or toroidal resonators, often fabricated in silica and smoothed by thermal reflow (for example by laser heating) which minimizes sidewall roughness. These structures reached high quality factors of \SI{1e8}{} very early on \cite{armaniUltrahighQToroidMicrocavity2003} and recently came close to the material limit of fused silica at telecom and near IR wavelength (over \SI{3e9}{}) \cite{shiOnchipUltrahighQOptical2025}. This manufacturing technique, however, is limited to only a few amorphous materials (mostly fused silica) and is difficult to integrate. The exact shape and size of such resonators is also not very controlled due to the reflow process and thus offers limited flexibility for photonic circuits.

Most development nowadays focuses on the ring resonators based on waveguides with higher refractive index than the substrate material (for example lithium niobate on silicon oxide) which can be fabricated with (semi-)standardized lithographical methods. This allows one to choose from a wider range of materials, in parcticular single crystalline ones which often allow for second order nonlinearities. There are roughly two classes of waveguides.
The first one has a very wide (a few micrometers) but thin (about 100 nanometers) high index core. This reduces the interaction of the modes with the etched sidewalls and hence minimizes scattering losses. The drawback is that the field overlap of the mode with the waveguide material is small and the confinement is weak which reduces the potential for nonlinear experiments and requires large bending radii to avoid radiative losses and intermodal coupling.

The second type, ridge waveguides, does not have this drawback. Here, the light is largely guided in the material of interest within a typical cross-section of a few hundreds of nanometers in both dimensions. This allows for strong confinement of the modes and small bending radii given the refractive index contrast of the core and cladding is large enough. The resulting small mode volumes are ideal for nonlinear interactions, while the high sensitivity of the propagation constant to the waveguide dimensions allows for efficient dispersion engineering. The drawback of this configuration is a strong sensitivity to sidewall roughness which requires highly optimized and developed manufacturing capabilities. However, this engineering challenge has been addressed in the past decade, and there has been tremendous progress. In the following we give an overview for the most promising materials, such as silica \ce{SiO2}, silicon nitride \ce{Si3N4} \cite{enjavi_silicon_2025, blumenthalSiliconNitrideSilicon2018, buzaverovSiliconNitrideIntegrated2024}, aluminium nitride \ce{AlN} \cite{li_aluminium_2021, liu_aluminum_2023}, lithium niobate \ce{LiNbO3} \cite{labbeThinfilmLithiumNiobate2025, liuRecentAdvancesLithium2025, zhu2021LNOfilms, qiIntegratedLithiumNiobate2020, xieRecentDevelopmentIntegrated2024}, and lithium tantalate \ce{LiTaO3}, as well as a few others. The importance of these materials extends beyond achieving high optical quality factors and low propagation losses. They provide access to additional functionalities that can substantially expand the capabilities of resonator-based microphotonic systems. In particular, AlN, SiC, and GaN exhibit appreciable second-order optical nonlinearities, enabling efficient second-harmonic generation, parametric frequency conversion, and other three-wave mixing processes. Their electro-optic and, in some cases, piezoelectric properties also facilitate direct electrical control of optical resonances and coupling between optical, microwave, and acoustic fields. AlGaAs, in turn, offers exceptionally strong second- and third-order optical nonlinearities, high refractive index, and compatibility with semiconductor optoelectronic technologies. These properties are particularly attractive for realizing compact, low-threshold nonlinear optical devices, including parametric oscillators, frequency converters, and Kerr frequency-comb generators.

A comparison of selected experimentally demonstrated intrinsic quality factors, key material properties, technological maturity, and representative applications of these integrated resonator platforms is provided in Table~\ref{tab:integrated-materials}.

\clearpage
\onecolumn
\begin{landscape}
\thispagestyle{plain}
\begin{center}
\begin{minipage}{\linewidth}
\centering
\begin{threeparttable}
\captionof{table}{Representative material properties and demonstrated applications of integrated WGM and microring platforms. The listed values are intended to provide an approximate comparison rather than a definitive ranking, as reported material parameters can vary substantially across the literature, particularly for nonlinear and electro-optic coefficients. Such variations arise from differences in material composition, crystal quality, measurement wavelength, and experimental methodology. For the optical quality factors, we selected high experimentally demonstrated intrinsic values in integrated resonators, favoring strongly confined geometries where available rather than weakly confined structures optimized primarily for ultrahigh $Q$. The nonlinear and electro-optic coefficients represent material properties rather than effective interaction strengths in particular devices, while the transparency windows refer to bulk materials and do not necessarily reflect propagation losses in integrated waveguides.}
\label{tab:integrated-materials}
\renewcommand{\arraystretch}{1.08}
\setlength{\tabcolsep}{2.1pt}
\footnotesize
\begin{tabularx}{\linewidth}{@{}l C{30mm} C{28mm} C{29mm} C{28mm} C{33mm} C{20mm} L@{}}
\toprule
\textbf{Material} & \textbf{selected $\boldsymbol{\Qi}$} \textbf{(demonstrated)} & \textbf{$\boldsymbol{d_{ij}}$} \textbf{(pm/V)} & \textbf{Pockels $\boldsymbol{r_{ij}}$} \textbf{(pm/V)} & \textbf{$\boldsymbol{n_2}$} \textbf{($10^{-19}$ m$^2$/W)} & \textbf{Transparency} \textbf{($\boldsymbol\mu$m)} & \textbf{PIC maturity} & \textbf{Selected demonstrated applications} \\
\midrule
\textbf{SiO$_2$} & \(\, 1.1\times10^{9}\,\)\,\cite{wu_greater_2020} & --- & --- & \(\, 0.22\,\)\,\cite{schiekNonlinearRefractiveIndex2023} & \(\, \sim0.21\text{--}3.5\,\)\,\cite{bederTransmissivityAbsorptionFused1971} & Established & Kerr combs; Brillouin lasers \\
\addlinespace[2pt]
\textbf{Si$_3$N$_4$} & \(\, 3.7\times10^{7}\,\)\,\cite{Ji2017Si3N4} & --- & --- & \(\, \sim2.4\,\)\,\cite{ikedaThermalKerrNonlinear2008} & \(\, \sim0.4\text{--}6.7\,\)\,\cite{beliaev2022, kischkatMidinfraredOpticalProperties2012} & Established & Kerr combs; SFWM photon pairs; squeezing \\
\addlinespace[3pt]
\textbf{LiNbO$_3$ (LN)} & \(\, 2.9\times10^{7}\,\)\,\cite{zhu2024highQ} & \(\, 25.2\ (|d_{33}|)\,\)\,\cite{shojiAbsoluteScaleSecondorder1997} & \(\, 28.8\ (r_{33}^{S})\,\)\,\cite{weisLithiumNiobateSummary1985} & \(\, \sim1.8\,\)\,\cite{zhu2021LNOfilms} & \(\, \sim0.35\text{--}5\,\)\,\cite{leidingerComparativeStudyThree2015} & Scaling & EO combs; SHG/SPDC; microwave--optical conversion \\
 & & \(\, 4.6\ (|d_{31}|)\,\)\,\cite{shojiAbsoluteScaleSecondorder1997} & \(\, 7.7\ (r_{13}^{S})\,\)\,\cite{weisLithiumNiobateSummary1985} & & &  & \\
 & & \(\, 2.1\ (d_{22})\,\)\,\cite{millerDependenceSecondHarmonicGenerationCoefficients1971} & \(\, 18.2\ (r_{51}^{S})\,\)\,\cite{kaminowCrystallographicElectroopticProperties1980} & & &  & \\
 & & & \(\, 3.4\ (r_{22}^{S})\,\)\,\cite{weisLithiumNiobateSummary1985} & & &  & \\
\addlinespace[3pt]
\textbf{LiTaO$_3$ (LT)} & \(\, 2.3\times10^{7}\,\)\,\cite{heLithiumTantalateMicroring2025} & \(\, 13.8\ (|d_{33}|)\,\)\,\cite{shojiAbsoluteScaleSecondorder1997} & \(\, 30.5\ (r_{33}^{T})\,\)\,\cite{onukiInterferometricMethodMeasuring1972} & \(\, 1.7(o) - 3.0(eo)\,\)\,\cite{ashiharaNonlinearRefractionFemtosecond2003} & \(\, \sim0.28\text{--}5.5\,\)\,\cite{humOpticalPropertiesFerroelectric2007} & Scaling & EO combs; Kerr combs; EO modulation \\
 & & \(\, 0.85\ (|d_{31}|)\,\)\,\cite{shojiAbsoluteScaleSecondorder1997} & \(\, 8.4\ (r_{13}^{T})\,\)\,\cite{onukiInterferometricMethodMeasuring1972} & & &  & \\
 & & $>1.5\ (|d_{22}|)$\,\cite{Dolev2009LTa}& \(\, 20\ (r_{51}^{S})\,\)\,\cite{zhu2021LNOfilms} & & &  & \\
\addlinespace[3pt]
\textbf{AlN} & \(\, 3.7\times10^{6}\,\)\,\cite{liuFundamentalLinewidthAlN2022} & \(\, 4.3\ (|d_{33}|)^{\dagger}\,\)\,\cite{majkicOpticalNonlinearElectrooptical2017} & \(\, 1.16\ (r_{33}^{T})\,\)\,\cite{majkicOpticalNonlinearElectrooptical2017} & \(\, \sim3.5\,\)\,\cite{liuIntegratedHighQCrystalline2018} & \(\, \sim0.20\text{--}10\,\)\,\cite{beliaevThicknessdependentOpticalProperties2021,gadenneOpticalPropertiesAlN1998} & Developing & Kerr combs; SHG; OPO \\
 & & \(\, \sim0.1\ (|d_{31}|)^{\dagger}\,\)\,\cite{majkicOpticalNonlinearElectrooptical2017} & \(\, 0.11\ (r_{13}^{T})\,\)\,\cite{majkicOpticalNonlinearElectrooptical2017} & & &  & \\
\addlinespace[3pt]
\textbf{4H--SiC} & \(\, 5.8\times10^{6}\,\)\,\cite{zhengOctaveSpanningDeterministicSingle2026} & \(\, -11.7\ (d_{33})\,\)\,\cite{satoAccurateMeasurementsSecondorder2009} &\(\, \lesssim 0.03\ (|r_{33}|)\,\)\,\cite{wangInvestigationElectroopticEffect2023}  & \(\, 6.0\,\)\,\cite{zheng4HSiCMicroringResonators2019} & \(\, \sim0.37\text{--}5.6\,\)\,\cite{wang4HSiCNewNonlinear2013} & Developing & Kerr combs; harmonic generation; Raman lasing \\
 & & \(\, 6.5\ (d_{31})\,\)\,\cite{satoAccurateMeasurementsSecondorder2009} & \(\, 0.3\text{--}0.7\ (|r_{13}|)\,\)\,\cite{wangInvestigationElectroopticEffect2023}& & &  & \\
 & & \(\, 6.7\ (d_{15})\,\)\,\cite{satoAccurateMeasurementsSecondorder2009} & & & &  & \\
\addlinespace[3pt]
\textbf{AlGaAs} & \(\, 3.5\times10^{6}\,\)\,\cite{xieUltrahighQAlGaAsoninsulatorMicroresonators2020a} & \(\, \sim105\ (d_{14})^{\ddagger}\,\)\,\cite{ulsigEfficientWidelyTunable2024} & \(\, \sim -1.43\ (r_{41})^{\ddagger}\,\)\,\cite{glickLinearElectroopticEffect1988} & \(\, \lesssim 230^{\ddagger}\,\)\,\cite{wathenEfficientContinuouswaveFourwave2014} & \(\, \sim0.7\text{--}17^{\ddagger}\,\)\,\cite{mobiniAlGaAsNonlinearIntegrated2022a} & Developing & Kerr combs; SPDC; SHG/SFG \\
\addlinespace[3pt]
\textbf{GaN} & \(\, 2.5\times10^{6}\,\)\,\cite{zhengIntegratedGalliumNitride2022a} & \(\, -3.8\ (d_{33})\,\)\,\cite{abeAccurateMeasurementQuadratic2010} & \(\, 1.60\ (r_{33})\,\)\,\cite{cuniot-ponsardElectroopticConversepiezoelectricProperties2014} & \(\, 14\,\)\,\cite{zhengIntegratedGalliumNitride2022a} & \(\, \sim0.37\text{--}7\,\)\,\cite{manasrehOpticalAbsorptionBand1996,welnaTransparencyGaNSubstrates2012} & Emerging & SHG; waveguide-coupled WGM emission; Kerr combs \\
 & & \(\, 2.5\ (d_{31})\,\)\,\cite{abeAccurateMeasurementQuadratic2010} & \(\, 1.00\ (r_{13})\,\)\,\cite{cuniot-ponsardElectroopticConversepiezoelectricProperties2014} & & &  & \\
 & & \(\, 2.5\ (d_{15})\,\)\,\cite{abeAccurateMeasurementQuadratic2010} & & & &  & \\
\bottomrule
\end{tabularx}
\begin{tablenotes}[flushleft]
\scriptsize
\item \textbf{Notes.} PIC maturity is qualitative. Dashes for amorphous SiO$_2$ and Si$_3$N$_4$ denote absent bulk electric-dipole $\chi^{(2)}$ and Pockels responses, not engineered or interface effects. The SiC entries refer specifically to 4H--SiC. Tensor signs follow the cited conventions; the LN/LT $d$ coefficients from Shoji et al. are shown as magnitudes. Superscripts $T$ and $S$ denote mechanically unclamped and clamped Pockels coefficients, respectively. Most cited EO values were measured at low frequencies; the 4H--SiC $r_{13}$ range was inferred at 10\,kHz. The LN, LT, 4H--SiC and GaN $d$ values were measured at 1064\,nm. The GaN $r$ values were measured in an epitaxial film. For LT, $|d_{22}|$ is a lower bound for MgO-doped stoichiometric material, and $n_2$ was measured near 800\,nm.
\item[$\dagger$] The AlN $d$ values are measured magnitudes at 1030\,nm, with $|d_{33}|=4.3\pm0.3$\,pm/V and $|d_{31}|=|d_{33}|/(45\pm5)$. The direct AlN result \cite{majkicOpticalNonlinearElectrooptical2017} differs from other tabulations \cite{zhu2021LNOfilms}.
\item[$\ddagger$] AlGaAs properties vary strongly with alloy fraction and wavelength. The quoted $r_{41}$ was measured in Al$_{0.17}$Ga$_{0.83}$As at 1152.3\,nm; the $n_2$ bound applies to the alloys studied near 1550\,nm. The transparency range is an approximate literature estimate. $d_{14}$ follows the $d=\chi^{(2)}/2$ convention.
\end{tablenotes}
\end{threeparttable}
\end{minipage}
\end{center}
\end{landscape}
\twocolumn

\subsection{Silica}

Silica resonators have been traditionally shaped in the reflow process \cite{armaniUltrahighQToroidMicrocavity2003,shiOnchipUltrahighQOptical2025}. The micro-toroids and micro-spheres fabricated in this way feature high $Q$-factors and are still frequently used for nonlinear optics research. For example, coating such resonators with indium tin oxide nanoparticles was shown to significantly increase the third-harmonic generation efficiency \cite{pampel2020THG}.

\begin{figure}[htbp]
\centering\includegraphics[width=8cm]{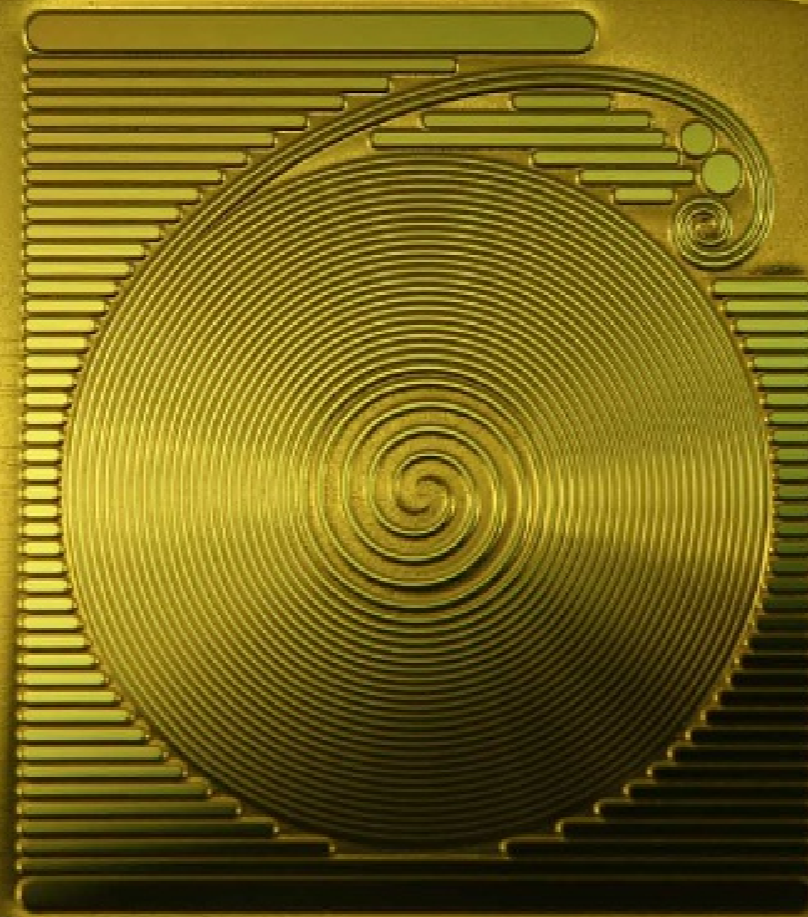}
\caption{A silica-on-silicon spiral resonator has a large length (1.2 m) and small footprint (2 cm wide). Reproduced from \cite{lee_spiral_2013}.
%This article is licensed under a Creative Commons 3.0 License, which permits the reproduction
}
\label{fig:lee_spiral_2013Fig1}
\end{figure}

A more novel and PIC-friendly approach developed by Lee et al.~\cite{lee_chemically_2012} is based on a direct chemical undercutting of lithographically defined preforms. It leads to an on-chip Q factor of 875 million. The optimization of this process by Wu et al.~\cite{wu_greater_2020} boosted the Q factor to 1.1 billion, which is close to the theoretical limit of silica at 2.5 billion \cite{lee_chemically_2012}. Various shapes and sizes of on-chip resonators can be fabricated in this way. The ridge-shaped ring resonators with $Q>2\times 10^8$ have been successfully used for generating solitons and as Brillouin lasers \cite{yang_bridging_2018}. To increase the resonator size on the same chip footprint, these methods were used to create spiral resonators, e.g. silica-on-silicon \cite{lee_spiral_2013} as well as silicon-on-insulator \cite{igarashi2023SS}. An example of such a spiral is shown in Fig.~\ref{fig:lee_spiral_2013Fig1}.

An effective and versatile method of frequency-tuning of silica PIC resonators with an auxiliary laser used as a heater was demonstrated by \cite{koehler2018Si}. This technique should be applicable to other resonator materials.

\subsection{Lithium niobate and lithium tantalate}

Lithium niobate (LN) and lithium tantalate (LT) are canonical materials for nonlinear optics and have been used for decades in bulk WGM resonators. Both have low optical loss from the visible into the infrared regime, a high enough refractive index ($>2$) for mode confinement in a fused silica cladding, and a strong second order nonlinear coefficients of up to \SI{30}{\pico\meter\per\volt} depending on polarisation and crystal orientation. They are well established in the industry and therefore high quality single crystal wafers are commercially available for both materials.
Fundamental for scalable LN and LT photonic chip manufacturing, however, has been the availability of single crystalline thin films of these materials on silicon or silica. A large review by Zhu et al. \cite{zhu2021LNOfilms} is dedicated to those thin film wafers and their applications. A decisive advance in this area was achieved by using the smart-cut technology \cite{poberaj_lithium_2012}, where a sub-micrometer thin layer from a single bulk crystal of LN or LT is transferred onto a silicon oxide substrate on a wafer scale and subsequently polished to remove surface damage and roughness. Commercial availability of such LN on insulator (LNOI) wafers (for example NanoLN) has enabled rapid development of low-loss on chip LN waveguides. From here on, typical lithographic methods can be used to create photonic structures. Already in 2014 Wang et al.~\cite{wangIntegratedHighQuality2014} demonstrated that on-chip LN microdisks (not a ridge waveguide ring yet) can be manufactured with e-beam lithography and a combination of dry and wet etching reaching Q factors of \SI{1e5}{}. Those disks had to be underetched to allow for efficient coupling to a tapered fiber. In 2017, the same group improved their fabrication process and design significantly and presented an integrated lithium niobate ring resonator with a Q factor of \SI{1e7}{} in the telecom regime coupled to an on-chip ridge waveguide \cite{zhang_monolithic_2017}. More recently, they also showed that the Q factor can be increased to \SI{2.9e7}{} by further optimizing the etching process and changing the design from a ring to a racetrack geometry which reduces sidewall scattering \cite{zhu2024highQ}.
Remarkably, it was shown that LNOI could in principle reach the same low absorption as bulk LN wafers. While the smart-cut process introduces absorptive defects, they can be largely removed by thermal treatment in an oxygen atmosphere. It was demonstrated that this could lead to absorption limited Q factors of \SI{1.6e8}{} given that the sidewall scattering can be further reduced \cite{shams-ansariReducedMaterialLoss2022}. 
A non-lithographic method for making high-$Q$ LN ring resonators has been presented by Wu et al. \cite{wuLithiumNiobateMicrodisk2018} and Gao et al. \cite{gaoLithiumNiobateMicroring2021}: to avoid ion-induced damage, they fabricate their own LNOI wafer by bonding and polishing (instead of using smart-cut). Then they use a femtosecond laser patterned chromium hard mask on an LNOI wafer for selective chemo-mechanical polishing. The resulting disk and ring structures have low sidewall roughness and no ion-induced damage and therefore reach intrinsic $Q$ factors up to $10^8$. This method offers less flexibility for the layout than standard lithographic processes.

More recently, people started to shift attention towards lithium tantalate (LT). Its major properties (nonlinearity, transparency, refractive index) are similar to those of LN but slightly more beneficial: LT has slightly lower optical and microwave losses, a slightly larger optical transparency window, and has a smaller birefringence than LN which leads to a reduced intermodal coupling between TE and TM modes in waveguide bends in x-cut geometries. The reason for this is that the TE mode (in-plane polarization) experiences a direction depended propagation constant which can match the TM mode in LN but not in LT where geometric dispersion is stronger than the birefringence. This was theoretically and experimentally confirmed \cite{panFundamentalModeHybridization2019,wang2024LTA}.
On top of that, LT is a more established material in the industry since it is used in 5G filters indicating a shorter pathway towards mass production. 
The manufacturing processes of LT integrated chips are largely the same as those of LN \cite{wang2024LTA} and ring resonators with an intrinsic Q of \SI{1e7}{} have been demonstrated \cite{heLithiumTantalateMicroring2025}.
So far, most LN and LT on-chip devices have been fabricated using e-beam lithography which is not well suited for scaling. However, it was recently shown that good quality, very densely packed chips can also be fabricated using UV stepper lithography in LN \cite{liHighDensityLithium2023} and LT \cite{wang2024LTA}. Both works provide a statistical analysis of their on chip resonators and modes showing an average intrinsic Q factor of approximately \SI{5e6}{} while a few modes also reach the \SI{1e7}{} regime.

In both, LN and LT integrated resonators, impressive nonlinear experiments have been demonstrated. Electro-optic frequency comb generation \cite{caiHighQIntegratedLithium2024a, zhangUltrabroadbandIntegratedElectrooptic2025a}, Kerr comb generation \cite{cai_soliton_2025}, and based on that the realization of synthetic optical crystals \cite{huRealizationHighdimensionalFrequency2020, huMirrorinducedReflectionFrequency2022, javidChipscaleSimulationsQuantumcorrelated2023} and electro-optic conversion for radiometry \cite{zhangIntegratedPhotonicMillimeterwave2025a} to name only a few examples.

\subsection{Silicon nitride} \label{sec:Si3N4}

Silicon nitride  (\ce{Si3N4}) is currently the most established material for CMOS compatible photonic chip manufacturing. On chip, it is usually amorphous and hence limited to third order nonlinearities but has otherwise attractive properties: a refractive index around 2 allows for good confinement and low (radiative) bending losses on waveguides. It has a low absorption over a large transparency window from about \SIrange{400}{2350}{\nano\meter} \cite{blumenthalSiliconNitrideSilicon2018} and does not show two photon absorption in the telecom band \cite{kruckelOpticalBandgapEngineering2017}. Due to its thermal stability and high optical damage threshold it can handle high power and is therefore suited for third order nonlinear optics experiments, which is backed by its reasonably large Kerr nonlinearity (10 times larger than fused silica \cite{ikedaThermalKerrNonlinear2008})\\
For manufacturing low loss waveguides and high Q ring resonators on chip, three major problems had to be solved over the last decade(s): sidewall scattering, stress induced cracking, and process induced absorptive impurities. An older solution to this was to make very thin and wide \ce{Si3N4} core waveguides (typically \SI{40}{nm} thickness and \SI{11}{\micro\metre} width) which enforces a large bending radius due to weak confinement. The intrinsic Q factors for this type of resonators reached almost \SI{1e8}{} quite early \cite{spencerIntegratedWaveguideCoupled2014a} and has been more recently pushed to \SI{4.22e8}{} \cite{puckett2021Si3N4} and \SI{7.2e8}{} \cite{liuUltralow0034DB2022}.

Improvements in the fabrication process lead to smoother sidewalls and top surfaces. By optimizing etch and hard mask and introducing chemo-mechanical polishing and multipass e-beam lithography higher Q factors of \SI{6.7e7}{} in rings made of ridge waveguides have been achieved \cite{Ji2017Si3N4}. This trend was followed by introducing the so called photonic Damascene process \cite{pfeiffer_photonic_2018,pfeifferUltrasmoothSiliconNitride2018} where a negative of the waveguide structure is lithographically written into a \ce{SiO2} layer which can be annealed to reduce roughness. Afterwards \ce{Si3N4} is deposited into the trenches, polished flat, annealed to drive out hydrogen impurities, cladded with \ce{SiO2}, and annealed a third time. This process was demonstrated to work with DUV stepper technology and produce realiably intrinsic $Q$ factors above \SI{1e7}{} \cite{liuHighyieldWaferscaleFabrication2021}. Unfortunately, the need for polishing and annealing temperatures above \SI{1000}{\degreeCelsius} is detrimental for material flexibility and large scale manufacturing. It was shown  that polishing can be skipped and annealing temperatures can be reduced to \SI{250}{\degreeCelsius} if the \ce{Si3N4} deposition is improved by using plasma assisted chemical vapor deposition (CVD) instead of the low pressure CVD, and using deuterated process gasses to reduce absorption by defects \cite{bose_anneal-free_2024}. Using this more foundry friendly method loaded Q factors of \SI{4e6}{} have been achieved. Even higher values $Q\approx 10^7$ were achieved by room-temperature reactive sputtering \cite{zhang2024Si3N4}.

The major use case for \ce{Si3N4} on-chip resonators is the generation of Kerr combs, which was demonstrated already in 2010 by Levy et al. \cite{Levy10comb} as a multi-frequency OPO. Shortly after, it was demonstrated that such an OPO produces indeed equidistant (comb) lines \cite{fosterSiliconbasedMonolithicOptical2011} which can span an octave \cite{Okawachi11octave_comb}. A few years later, dual comb generation and spectroscopy was conducted on a single chip with a single pump laser \cite{duttOnchipDualcombSource2018}. Further steps towards a full integration have been undertaken by Xiang et al. \cite{xiangLaserSolitonMicrocombs2021}. They fabricated a hybrid chip where a \ce{Si3N4} microring is directly coupled to a heterogeneously integrated InP/Si semiconductor laser, both sitting on the same monolithic silicon chip. The laser is injection-locked to the resonator and serves as the pump for comb generation.
A complementary direction is the system-level integration of \ce{Si3N4} Kerr-comb sources with active components. Sun et al.~\cite{sunChipintegratedCombbasedMicrowave2025} demonstrated a Kerr comb based microwave oscillator on a hybrid chip by combining a DFB laser, a \ce{Si3N4} Kerr microresonator, and a high-speed photodetector.

\subsection{Aluminum nitride and other unconventional materials}

Aluminum nitride (AlN) has emerged as a highly versatile platform for integrated WGM resonators over the past decade. It has an ultra-wide transparency window (from deep ultraviolet \SI{200}{\nano\meter} to mid-infrared \SI{13.6}{\micro\meter} \cite{majkicOpticalPropertiesAluminum2015}), an exceptionally high thermal conductivity for an insulator (\SI{320}{\watt\per\meter\per\kelvin} \cite{taylorPropertiesAluminumNitride1960}) reasonable second order optical nonlinearities \cite{majkicOpticalNonlinearElectrooptical2017}, and strong piezoelectricity \cite{tsubouchiZeroTemperatureCoefficientSAWDevices1985}. As comprehensively surveyed in the 2023 review by Liu et al.\ \cite{liu_aluminum_2023} fabrication started with sputtering thin layers of AlN on silicon (and other substrate materials) with a low process temperature of \SI{350}{\degreeCelsius} and subsequent e-beam lithography \cite{xiongAluminumNitrideNew2012}. This results in poly-crystalline films with a strong c-axis orientation allowing second order nonlinear processes. This process is simple and CMOS compatible but the poly-cystallinity leads to significant surface roughness of the waveguides which limits the Q factor of ring resonators to the $10^5$ regime at telecom wavelengths \cite{bruch2018_SHG, surya2018THG}.
It was demonstrated by Liu et al.\ \cite{liuAluminumNitrideonsapphirePlatform2017} that high quality single-crystalline AlN films can be grown on a sapphire substrate via metal-organic chemical vapor deposition and subsequent annealing. The substrate and the high temperatures ($> \SI{1000}{\degreeCelsius}$) make the process less foundry friendly, but improve optical properties significantly. Quality factors well exceeding $10^6$ at the telecom wavelengths \cite{sunUltrahighMicroringResonators2019, dingHighQAlNMicroresonators2026} and at \SI{780}{\nano\meter} \cite{dingHighQAlNMicroresonators2026} have been reported. In the visible (\SI{638}{\nano\meter}) and UV (\SI{369.5}{\nano\meter}) ranges the $Q$-factors of respectively \SI{1.7e5}{} and \SI{2e4}{} have been demonstrated in nanocrystals \cite{luAluminumNitrideIntegrated2018}. With actual single-crystalline AlN, the Q factor at \SI{390}{\nano\meter} was pushed up to \SI{2.1e5}{} \cite{liuUltrahighQUVMicroring2018}.

Integrated AlN resonators are often used to generate Kerr combs, but since single crystalline thin films are available efficient second harmonic generation \cite{guoSecondharmonicGenerationAluminum2016a,bruch2018_SHG} and parametric down conversion (OPO) \cite{bruch2019microring} have been demonstrated as well. A combination of both effects enabled an octave spanning Kerr comb which was self referenced with frequency doubling of the pump frequency on the same chip \cite{liuAluminumNitrideNanophotonics2021}.

%\subsection{Other materials}

Besides the materials discussed above, there are several other perhaps less popular but nonetheless potentially important materials that are investigated as relevant candidates for integrated resonators. 

Silicon carbide (\ce{SiC}) has attracted strong interest because it combines a wide bandgap with a strong optical nonlinearity approaching that of lithium niobate. Also it has defects which are interesting for certain quantum applications such as single photon sources. A demonstration of reasonable $Q$ factors up to \SI{7e6}{} \cite{wang2021SiC} nonlinear optical processes in SiC-on-insulator microrings \cite{zheng2019SiC,wang2021SiC,zhang2024SiC} and its use as a single photon source \cite{lohrmannReviewSinglePhoton2017} suggests that it might become a versatile integrated resonator platform rather than a niche material. 

%Diamond can host several optically addressable spin defects (color centers), most prominently, NV centers, which are interesting for quantum memories, magnetic sensing and other quantum applications. In parallel it is also a suited material for optical resonators as it has a wide transparency window, high refractive index, exceptional thermal conductivity, and high optical damage threshold \cite{zaloginaEngineeringQuantumLight2026}. 
%Diamond resonator technology \cite{dory2019diamond,regan2019diamond,mi2020diamond,shandilya2022diamond_rev} remains less mature from a fabrication and integration perspective, but its exceptional transparency window, thermal conductivity, damage threshold, and compatibility with quantum-defect photonics make it an interesting candidate. 
%Early high-$Q$ work \cite{hausmannIntegratedHighQualityFactor2013} but difficult fabrication. Field remains a Niche since people focus on Purcell enhancement which is currently better in other types of resonators (nanobeams, phtonic crystals, ...)

%Gallium phosphide (GaP) is particularly attractive for visible and telecom nonlinear photonics because of its high refractive index, strong optical nonlinearity, and reduced two-photon absorption at telecom wavelength. Recent works indicate rapidly growing momentum for GaP-on-insulator devices \cite{schneider_optomechanics_2019,anthur2021GaP_SHG,wilson2020GaPcomb}. 

Aluminum gallium arsenide (\ce{AlGaAs}) is attractive since it has one of the largest optical nonlinearities ($\chi^{(2)}_{xyz} \approx$ 238 pm/V and $n_2\approx 2.6\times10^{-17}$ m$^2$/W in the telecom band, \cite{talenti2025GaAs}) among the materials used for integrated photonics. Furthermore, alloy engineering by changing the ratio between Ga and Al allows to tailor the refractive index (\SIrange{2.9}{3.4}{}) and the dispersion of the material \cite{mobiniAlGaAsNonlinearIntegrated2022a}. 
Together with reasonably high $Q$ factors in the \SI{1e6}{} regime \cite{xieUltrahighQAlGaAsoninsulatorMicroresonators2020a}, these properties enable very compact nonlinear devices, where both the strong Kerr nonlinearity and the second-order susceptibility can be exploited. As a result, \ce{AlGaAs} has been used for low-threshold Kerr frequency comb generation \cite{changUltraefficientFrequencyComb2020a}, bright and high-purity quantum-light sources from spontaneous parametric down conversion \cite{autebertIntegratedAlGaAsSource2016,steinerUltrabrightEntangledPhotonPairGeneration2021}, as well as other $\chi^{(2)}$ based processes such as second-harmonic generation \cite{maySecondharmonicGenerationAlGaAsoninsulator2019}, or difference-frequency generation \cite{babouxN2023}. In addition, its III--V semiconductor nature makes the platform attractive for future integration with active components such as electrically injected lasers, quantum-dot emitters, modulators, and on-chip detectors \cite{babouxN2023}.

Finally, gallium nitride (\ce{GaN}) offers a wide bandgap, good robustness, and attractive operation in the visible and near-UV, which may make it increasingly relevant for shorter-wavelength nonlinear and quantum photonics \cite{roland2016SHG_GaN,to2020GaN}.

Despite their considerable potential, the implementation of AlN, SiC, AlGaAs, and GaN in microring resonators faces several unresolved materials and technological challenges. A major issue is achieving simultaneously high intrinsic quality factors and reproducible device characteristics using scalable fabrication processes. The results vary significantly from experiment to experiment. Sidewall roughness, crystalline defects, grain boundaries, residual stress, and material inhomogeneity can significantly degrade resonator performance making reliable wafer-scale production rather challenging. Additional difficulties arise from thermal effects, free-carrier absorption, photorefractive effects, and optical damage, whose relative importance varies substantially among materials and operating wavelengths. %For practical applications, it is also necessary to establish reliable electrical and thermal interfaces, efficient coupling to external waveguides and optical sources, and long-term stability under relatively high optical power and environmental variations. 
This leaves an important open question is whether these materials can deliver a sufficiently compelling combination of optical performance, multifunctionality, and fabrication reproducibility, to justify their adoption over the increasingly mature SiN technologies.

\section{Second-order nonlinear processes}\label{sec:chi2}

Second-order nonlinear processes are enabled by optical quadratic nonlinearity $\chi^{(2)}$ or electro-optic nonlinearity $r$, available in non-centrosymmetric crystals. These processes can be greatly enhanced in high-$Q$ crystalline resonators. Previously we discussed the resonator-based realizations of electro-optical conversion, second-harmonic generation, parametric down conversion and of other second-order processes. In the following we review the relatively recent progress in these directions.

\subsection{Microwave detection and conversion}\label{sec:MW}
WGM resonators enable coupling the electronic side of the electromagnetic spectrum to the optical domain via electro-optic and piezoelectric effects. Minet et al.~\cite{minet2020sweep, minet_electro-optic_2021} used an adiabatic frequency conversion technique based on Pockels effect to change the frequency of light stored inside a resonator as a function of the voltage applied to the resonator. They validate the coherence and the application of their adiabatic frequency shifter using frequency-modulated continuous-wave Light Detection and Ranging (LiDAR)~\cite{mrokon_continuous_2024}. They also explored the piezoelectric response of BMF to tune the optical resonances of a WGM resonator~\cite{mrokon_electrically_2025}. Recently the authors also identified that mechanical resonances in such systems limit the linearity of such frequency conversion~\cite{mrokon_influence_2026}.

The more popular application of electro-optic frequency conversion in WGM resonators is for the upconversion of microwave and terahertz frequencies with two main fields driving progress on this front: satellite-based radiometry and quantum transduction of microwave qubits. 

The advantage of the WGM resonator platform having a compact footprint makes it suitable for implementation on modern smaller cheaper satellite technologies such as CubeSats where the development of bespoke receivers for space-based measurements is not uncommon. The sensitivity and noise limits of such a photonics-based receiver is compared with commercial microwave detectors (based on low-noise amplifiers) in \cite{santamaria-botello_sensitivity_2020}. Abdalmalak et al.~\cite{abdalmalak2022} propose a hybrid integrated scheme with a monolithic disk resonator for the radiometric detection of the water line at \SI{183}{GHz}. Achieving phase-matching between the millimeter-wave signal and the optical signals in materials with high dispersion such as \LN{} is navigated by using a metal ring above the \LN{} disk whose distance can be varied to tune the exact frequency of the signal to be upconverted. This works around the tight constraints imposed on the fabrication of these delicate structures. Strekalov et al.~\cite{strekalov2022Receiver} discussed circumventing the strict phase-matching constraints of an all-resonant scheme altogether by using an external hollow metal waveguide cavity for the microwave mode at \SI{90}{GHz}. A metal needle is used to concentrate the microwave field to a spot on the rim of the \LT{} resonator. In a push towards upconversion of higher frequencies, Suresh et al.~\cite{suresh2025Upconversion} reported a scheme that allowed the upconversion of six signals centered at \SI{167}{GHz} separated by the optical FSR (\SI{\sim 10}{GHz}). The authors were able to show that the landscape of upconversion in such a system had additional non-resonant upconversion channels that could be used for nonlinear frequency conversion, albeit with a lower efficiency than the resonant channels. 

Even at lower GHz and microwave frequencies, WGM resonators have sustained strong interest over the past decade as a platform for upconversion. In \cite{lambert2020qute}, Lambert et al.\ discuss a range of implementations for coherent conversion between microwave and optical photons highlighting electro-optic interactions as one of the more promising approaches. In particular, the authors mention that for efficient microwave upconversion, triply-resonant schemes~\cite{soltani2017upconversion}, in which the two optical and the microwave signals all couple to resonant modes, are particularly attractive. This can be achieved by imposing that all three electromagnetic fields are resonantly confined in the dielectric WGM resonator disk as demonstrated in a \LN{} disk~\cite{santamaria_botello_sensitivity_2018} or where the microwave mode is confined by an external cavity as proposed in~\cite{javerzac-galy_-chip_2016, wang_high-efficiency_2022} and experimentally demonstrated in~\cite{rueda_efficient_2016}. 

Sekine et al.~\cite{sekine2026MWconvert} also present comparisons of microwave-optical transduction using the electro-optic effect as well as comparisons to other methods such as magneto-optic effects, MHz electro-optomechanical effects, and GHz optomechanical effects. There have been numerous demonstrations of electro-optic upconversion in both \LN{} platforms: bulk WGM resonators~\cite{hease_bidirectional_2020, khanna2026MWconvert} with the  photon number conversion efficiencies close to a percent, and TFLN~\cite{mckenna2020optica,holzgrafe2020optica,xu2021Bidirectional, hu_-chip_2021}. Xu et al~\cite{xu2021Bidirectional} show bidirectional frequency conversion with the photon number efficiency of 1\%. Warner et al~\cite{warner_coherent_2025} use coupled optical racetrack hybridized \LN{} resonators and demonstrate the conversion efficiency of 1.18\% . Singh et al~\cite{singh_efficient_2024} discuss incorporating a third resonator into the ``photonic molecule'' structure that was presented in~\cite{hu_-chip_2021} for tunable conversion from RF to optical carrier over two octave range of RF carrier frequency. M{\"o}hl et al.~\cite{mohl_bidirectional_2025} integrated a barium titanate optical resonator with a superconducting microwave resonator on chip to demonstrate the bidirectional conversion efficiencies of \SI{1e-6}{} with pulsed pumping. Sekine et al.~\cite{sekine_microwave--optical_2025} also discuss transduction mediated by antiferromagnetic magnons. A comparison of microwave to optical transduction efficiency observed in recent experiments is shown in Fig.~\ref{fig:efficiency_progress} together with a benchmark transduction efficiency of $5\times 10^{-7}$ achievable with a commercial electro-optic modulator operating at 6 GHz with $V_\pi=2.7$ V. We assumed a milliwatt-range optical power in the telecom range and used an efficiency expression that can be derived e.g. based on results by \cite{strekalov2022Receiver} or found in the explicit form in \cite{Peairs_transducers_2019}.

\input{transductionProgressFig}

The microwave to optics transduction can occur via either Stokes or anti-Stokes process corresponding to the frequencies subtraction or addition, respectively. The anti-Stokes transduction can be understood in the context of an optoelectronic cooling, when the optical pump photons and a microwave photons are annihilated and an anti-Stokes photon is created at a higher optical frequency. In this regime, the resonator mediates the conversion of microwave fluctuations into optical fluctuations. It can be shown that under the optimal conditions, the fluctuations of the output anti-Stokes field reproduce the fluctuations of the input microwave field and do not depend on the pump fluctuations. This enables the quantum-state-preserving frequency conversion wherein the quantum state initially encoded in the microwave photon can be transferred to the anti-Stokes optical photon.

The optimum conversion regime is achieved in the limit of a strong classical optical pump. It furthermore requires the critical coupling condition for the microwave mode, when the coupling rate to the external input channel equals the effective optoelectronic damping rate. The intrinsic microwave dissipation rate must remain much smaller than these rates, ensuring that the microwave excitation is mostly converted into the anti-Stokes optical mode and not dissipated internally. This balance enables near-unity conversion efficiency while preserving the quantum properties of the transferred state. Note that the microwave mode temperature is significantly lower than the ambient temperature since the thermal microwave photons are efficiently removed. 

The possibility of transferring a quantum state between signals with substantially different carrier frequencies, such as microwave and optical fields, suggests that the carrier frequency may be treated primarily as a classical phase evolution imposed on the underlying quantum state. In a phase-space representation, the quantum state of a microwave field undergoes rotation at the microwave carrier frequency. Following the state-transfer process, the representation of the quantum state in phase space retains the same structure, with the primary change being the rate of phase-space rotation corresponding to the new carrier frequency.

Measurement of the transferred state typically involves mixing the signal with a classical local oscillator. This process effectively removes the explicit dependence on the rapidly varying carrier phase and converts the quantum state information into a measurable quantity accessible through photodetection techniques.

An intriguing aspect of such frequency-conversion processes is that it can modify the power uncertainty associated with the quantum state. The conversion mechanism primarily acts on the photon-number uncertainty, which itself is independent of the carrier frequency. However, the corresponding energy (or power) uncertainty is directly proportional to the photon energy and therefore depends on the carrier frequency. Consequently, conservation of the photon energy requires an adjustment of the frequency uncertainty of the optical pump field involved in the conversion process.

%The unused pump photons can therefore serve as an energy reservoir or buffer that enables the conversion process while maintaining overall energy conservation. This simple consideration imposes an important constraint on the nature of allowed quantum transformations: a purely unitary process cannot arbitrarily change the carrier frequency of photons without affecting some aspect of their quantum state or the participating fields. A frequency conversion process must therefore involve a redistribution of quantum uncertainties, with the pump field providing the necessary degree of freedom to preserve fundamental conservation laws.

Algorithms for quantum state transfer from microwaves to optics were theoretically proposed by \cite{rueda2019_MW_ent} and experimentally tested by \cite{fan2018upconversion} and by \cite{sahu2022Quant, sahu2023MWent} who demonstrated low-noise microwave-to-optics conversion with relatively high efficiency, and then observed continuous-variable entanglement (two-mode squeezing) between optical and microwave fields. Some of the highest total conversion efficiencies achieved have been 2\% with an integrated AlN optical resonator~\cite{fan2018upconversion} and 15\% in a bulk \LN{} WGM resonator~\cite{sahu2022Quant}. Coherent control of superconducting microwave cavity coupled with an \LN{} optical resonator was demonstrated~\cite{qiu_coherent_2023}. Recently, \cite{arnold_all-optical_2025} presented qubit readout, and \cite{werner_electro-optic_2026} demonstrated the reconstruction of a non-classical state both in \LN{} resonators via electro-optic transduction.  

\subsection{Electro-optical combs}\label{sec:chi2combs}
The described microwave conversion schemes use typically sum-frequency generation where a microwave and a pump photon is destroyed to generate a signal photon. This avoids noise arising from parametric amplification in the difference frequency process. Typically, this can be achieved by pump detuning or by mode hybridization \cite{rueda_efficient_2016}. Both schemes also (intentionally) suppress cascading the parametric process over multiple free spectral ranges of the WGM resonator.
Such a cascade, however, can also be exploited: if the pump laser is locked on the centre of the pump mode and the microwave resonator is fed with a strong signal which matches the optical FSR, an electro-optic frequency comb can be generated by cascading the sum- and difference-frequency processes.
Compared to their Kerr counterparts, EO combs require less pump power and have significantly less complex dynamics making them easier to operate and more stable. Nevertheless, the field remains smaller compared to the field of Kerr combs, because engineering a triply resonant system (optical modes and a microwave mode) which is phase-matched is considerably more difficult and generally EO combs have less conversion efficiency and output power compared to Kerr combs.

We are aware of only a few publications where bulk resonators have been used to generate EO combs. An all-resonant approach was demonstrated with a LN resonator embedded in a 3D microwave cavity \cite{rueda2019EOcomb} with a span of \SI{35}{nm} with a microwave input power of \SI{20}{dBm}. This work was paralleled by an on-chip implementation (x-cut lithium niobate) which demonstrated a \SI{80}{nm} span with a similar microwave input power which was not even resonantly enhanced \cite{zhangBroadbandElectroopticFrequency2019a}. In the same work, they also demonstrated the generation of a tuneable dual comb by feeding two slightly detuned microwave signals to the same chip. A similar, but less efficient, experiment was later also done in a bulk x-cut LN resonator \cite{lambert2023DualComb}).
The conversion efficiency from the optical pump into the comb lines of this first integrated implementation was about \SI{0.3}{\%}. The reason for this low efficiency is that the pump mode becomes heavily undercoupled when the microwave tone is switched on, due to a large intrinsic nonlinear coupling rate. That problem was later solved by the same group based on their theoretical proposal \cite{buscainoDesignEfficientResonatorEnhanced2020} using an older idea \cite{kourogiCoupledcavityMonolithicOptical1996}: to compensate for the nonlinear change of coupling rates which limited the conversion efficiency they use a second ring resonator which is strongly overcoupled to the electro-optic racetrack resonator to feed the pump into the converter. When the microwave is switched on, the pump mode becomes critically coupled to the converter and the conversion efficiency increases. That approach increased the conversion efficiency to \SI{30}{\%} while a wider comb of \SI{132}{nm} was also demonstrated \cite{huHighefficiencyBroadbandOnchip2022}.

Overall, eo-combs tend to have less spectral width than Kerr based combs. It has been shown that this can be compensated by using multiple pump frequencies on a single converter to generate smaller combs at the desired spectral positions \cite{shams-ansariThinfilmLithiumniobateElectrooptic2022}. Using this technique they measured gas absorption and dispersion using actual dual comb spectroscopy. 
%Deeper integration on a monolithic chip with filtering and electro-optic modulation on-chip was shown by Wang et al.\ \cite{wang2019LNcomb}. They used integrated, dispersion-engineered high-Q lithium niobate microresonators to generate broadband Kerr frequency combs,  select a single comb line using an electrically controlled filter, and modulate the intensity of the selected line.

More recently, the performance of an integrated EO comb was dramatically increased by using lithium tantalate as material \cite{zhangUltrabroadbandIntegratedElectrooptic2025a}. The small birefringence suppresses intermodal coupling and allows the comb to reach wider spans. On top of that they also implemented a microwave resonator and optimized the single photon coupling rate reaching $g = 2 \pi\times \SI{2.2}{\kilo\hertz}$. They demonstrated a comb spanning \SI{450}{nm} with an on-chip power consumption of \SI{7}{W}, which includes the laser and the microwave driver.

\subsection{Second harmonic generation}\label{sec:SHG}

The second harmonic generation (SHG) continues to be one of the main motivations for studying the second-order nonlinear processes in microresonators. Our 2016 review \cite{Strekalov16rev} lists 13 experimental reports of the SHG in resonator made from a variety of nonlinear crystals, with the fundamental wavelength ranging from 490 to 1985 nm and the best reported slope efficiency reaching 4.5 mW$^{-1}$. There has been a significant progress in this field since then, mostly based on the thin-film on-chip micro-ring resonators. These resonators can have the z-cut or x-cut configurations, with the crystal optical axis parallel or perpendicular to the resonator axis of rotation, respectively. The phase matching in the z-cut resonators is achieved through the polarization dispersion, geometric dispersion of different order modes (especially significant in small resonators), or periodical poling. In the x-cut resonators, one typically relies on the cyclic phase matching \cite{Lin13cyclic} that is very broad but typically has a much lower efficiency.

The  SHG from near-infrared pump was observed in z-cut \cite{lin2016SHGx-cut,wolf2017cascaded,chen2018phasematch,luo2019PDC} and x-cut \cite{luo2017SHandPDC,trainor2018selective} lithium niobate microresonators, as well as in periodically poled z-cut thin-film lithium niobate resonators \cite{wolf2018QPM,lu2019PPLNring,chen2019SHG,Lu2020PPLN-SH}. The slope efficiency reported by \cite{Lu2020PPLN-SH} reaching 50 mW$^{-1}$, which is a more than a 10-fold improvement over the best efficiency reported in the review \cite{Strekalov16rev}. Notably, this result was achieved with a chip-integrated TFLN resonator and waveguide configuration, potentially allowing for scaling.

SHG was also observed in resonators made from other crystals. Its efficiency in AlN on-chip micro-rings \cite{bruch2018_SHG,bruch2019microring} reached 0.17 mW$^{-1}$.
Another realization of a precisely designed microring resonator made from the same material allowed for the SHG phase matching corresponding exactly to the two-photon transition of $^{85}$Rb (778.12 nm), with 0.018 mW$^{-1}$ efficiency \cite{surya2018SHGtoRb}. 

A low-efficiency SHG was observed in a z-cut GaN-on-Si micro-disk \cite{roland2016SHG_GaN}. Preliminary low-efficiency results were also reported for a GaAs microring resonator pumped at approximately 2 microns \cite{chang2019GaAs}, in which case the observed slope efficiency fell 6-7 orders of magnitude short of the theoretically predicted 10 mW$^{-1}$. 

The uniaxial cubic silicon carbide (3C-SiC) has the main quadratic nonlinearity tensor component that is off-diagonal: $\chi^{(2)}_{xyz}\approx 34$ pm/V. Utilizing this component for the SHG requires a special xy-cut orientation, with the optical axis lying in the resonator plane and both x and y axis making a 45$^\circ$ angle with this plane. This configuration with the cyclic phase matching was realized by \cite{li2024SiC} who observed the SHG from 787.7 nm pump as well as the parametric down conversion.

Centrosymmetric crystals such as silicon nitride do not possess an intrinsic quadratic nonlinearity, except when the inversion symmetry is broken, e.g. near the resonator boundary. Such a surface anisotropy allowed for a relatively efficient SHG, as well as the third harmonic generation via the frequency-adding process, in a silicon nitride microring \cite{Levy11SiNmicroring}. Recently, this cascaded process was expanded to also generate the fourth harmonic of the 1600 nm pump, arising from both the second harmonic frequency-doubling and a frequency sum of the fundamental and third-harmonic fields \cite{ghosh_fourth-harmonic_2026}. The Type-0 phase matching for the entire cascade was achieved through the geometrical dispersion of the high-order signal modes.

Besides the on-chip resonators, SHG was also observed in a stand-alone x-cut LB4 resonator \cite{tian2025SH_LB4} for three different pump wavelengths: 1560 nm, 795 nm, and 517 nm. The wide range of possible pump wavelengths in this case is an important advantage, particularly concerning the UV generation 517 nm $\rightarrow$ 258.5 nm. However the conversion efficiency for the latter wavelength reached only 0.002 mW$^{-1}$. 

\subsection{Parametric down conversion and optical parametric oscillators}\label{sec:PDC}

The progress in resonator-assisted parametric down conversion (PDC) and based on this process optical parametric oscillators (OPO) goes hand in hand with the SHG progress, as these are two sides of the same optical nonlinear process. It is therefore not surprising that some publications report both the SHG and PDC/OPO observations in the same resonator. We already mentioned works by \cite{luo2017SHandPDC} and by \cite{luo2019PDC} who observed PDC in the same lithium niobate micro-ring resonators they used for the SHG; and by \cite{li2024SiC} who observed both processes in a 3C-SiC micro-ring.

The studies that more specifically focus on the PDC characterize the OPOs in terms of their threshold power and wavelength tuning curves. For example, \cite{huang2025HOM} reported 1 $\mu$W threshold power for a non-degenerate 532 nm $\rightarrow$ 950 nm + 1209 nm Type-I OPO coupling three fundamental WGMs in a 1.4 mm diameter, z-cut \LN{} disk resonator. Type-I OPO was also implemented in a mm-sized WGM resonator with the intrinsic quality factor of $6\times 10^6$ made from a z-cut CdSiP$_2$ \cite{amiune_pump_2022}. This OPO had a threshold around 200 $\mu$W and was broadly tunable in the range from $\lambda_s=2708$ nm to $\lambda_i=3575$ nm. A resonator-based mid-infrared OPO was demonstrated by \cite{meisenheimer2017_IR_OPO}. In this case a 3.5-mm-diameter WGM resonator was fabricated from a silver gallium selenide (AgGaSe$_2$) crystal. It had the threshold in the mW range, and generating up to 0.8 mW of parametric light. Pumped at 1.57 $\mu$m wavelength, this OPO was tunable from 2 $\mu$m for the signal to 8 $\mu$m for the idler.

The high efficiency of quadratic nonlinear processes in the $\chi^{(2)}$ resonators allows for the self-pumped OPO approach, when the optical pump first generates the second harmonic light in the resonator, and then this light drives the OPO. In this case the degenerate OPO is phase matched by virtue of the SHG phase matching, but importantly, non-degenerate OPO can be phase-matched as well. Indeed, if the SHG phase matching allows the $(m_f,\omega_f)\rightarrow(2m_f,2\omega_f)$ conversion of the fundamental mode with the orbital number $m_f$ and eigenfrequency $\omega_f$, then the down-conversion  $(2m_f,2\omega_f)\rightarrow(m_f+M,\omega_f+\Omega_f)+(m_f-M,\omega_f-\Omega_f)$ is also allowed for the same mode families. This process will lead to formation of an optical comb around the fundamental wavelength $\omega_f$ whose number of lines $2M+1$ is initially determined by the dispersion of the free spectral range (FSR) $\Omega_f$ and the resonator linewidth.

The sum-frequency conversion between the lines of the fundamental comb are allowed by the same phase matching conditions. They should lead to building up the second harmonic comb $2\omega_f\pm\Omega_{SH}$. To a superficial look, this requires that the fundamental FSR $\Omega_f$ does not differ from the second-harmonic FSR $\Omega_{SH}$ by more than the resonator linewidth. This condition usually does not hold in mm-size WGM resonators, and yet the double comb can be generated, as experimentally demonstrated by \cite{hendry2020chi2comb,szabados2020_SHcomb,amiune2021_chi2comb} in 1064 nm $\leftrightarrow$ 532 nm pumped \LN{} resonators. It was also observed \cite{amiune2023Midinfrared} in a WGM crystalline resonator made from CdSiP$_2$, although in this case the initial nonlinear process was a degenerate PDC (1550 nm  $\rightarrow$ 3.1 $\mu$m) instead of the SHG. 

The emergence of the double-comb can be attributed to the modulation instability, see e.g.\ \cite{hansson13pra,ricciardi2020quad_comb}, associated with the cascaded frequency sum and difference generation. It leads to excitation of the WGMs symmetrically offset by a large $M$ from the pump and the corresponding second harmonic modes. In the frequency domain the modulation instability leads to a sparse comb with multiple FSR line spacing. In the temporal domain it corresponds to the Turing rolls of the pump and second-harmonic power. Reaching this regime requires the pump laser to exceed a certain power threshold, as well as to have a certain initial detuning from its mode. Further pump power increase leads to denser combs in the pump and second-harmonic domains. Their line spacing was shown \cite{szabados2020_SHcomb} to closely match the pump FSR, which means that the second-harmonic FSR is dynamically changed to match the fundamental FSR and enable the double-comb.

This comb-formation mechanism is very similar to that of Kerr-combs, but with the $\chi^{(3)}$ nonlinearity replaced by a cascaded $\chi^{(2)}$ nonlinearity. It similarly leads to formation of bright and dark solitons with the characteristic sech$^2$ envelopes, centered around the fundamental and second-harmonic wavelengths. Such $\chi^{(2)}$ solitons were theoretically discussed by \cite{buryak2002_chi2solitons} and experimentally demonstrated by \cite{bruch_pockels_2021} in an aluminum nitride microring resonator. A recent progress in theoretical understanding of these solitons in resonators is discussed in \cite{hansson2018_chi2solitons,
villois2019Soliton,podivilov2020_chi2soliton,puzyrev2022Ladder}.

A soliton comb in lithium niobate micro-ring resonators with simultaneous frequency doubling was reported by \cite{villois2019comb,he2019LNcomb,gong2020LNringComb} and theoretically discussed by \cite{smirnov2020_dblcomb}. The effect was also studied in lithium tantalate \cite{wang2024LTA},  GaP \cite{wilson2020GaPcomb} and silicon nitride \cite{xue2017SHcomb}.

%\subsection{Optical combs based on $\chi^{(2)}$}\label{sec:chi2combs}

\section{%Third-order nonlinear processes
Higher-order nonlinear processes}\label{sec:Third-order}

\subsection{Kerr combs and resonator solitons}\label{sec:comb}

%\paragraph{Recent reviews}
Kerr frequency combs produce a plethora of intriguing physical phenomena including soliton crystals \cite{cole2017SolitonCrystal,Karpov2019Dynamics} and breather solitons \cite{Lucas2017Breathing,yu2017Breather}. They also have multiple practical applications, e.g. in optical clocks \cite{newman2019clock}. The field is very broad and it was reviewed in literature over past decade \cite{pasquazi18pr,yao2024microcomb}. The photonic chip-based frequency combs were reviewed in \cite{gaeta2019CombRev,chang22np}.  Comprehensive reviews of SiC photonic \cite{zhang2024SiC} and other material platforms \cite{kovach20aop,yang2024comb} were recently published. We do not intend to provide a complete coverage of the Kerr frequency comb development field here. Instead, we focus on the recent progress in the areas prioritized in our original review \cite{Strekalov16rev}.

%\paragraph{Mode coupling for the group velocity dispersion optimization}
It has been shown that coupling between modes of a nonlinear optical microresonator can enable the generation of Kerr frequency combs \cite{Savchenkov12overmoded}. Nearly degenerate modes may interact as a result of geometric imperfections of the resonator and material inhomogeneities. Such interactions lead to shifts of the resonance frequencies and, consequently, to a modification of the effective group-velocity dispersion (GVD), thereby strongly influencing comb formation.

More recently, the role of coupling between clockwise and counter-clockwise propagating modes in ring microcavities has been investigated, revealing that this interaction can give rise to counter-propagating Kerr frequency combs \cite{fujii2017Comb}. These effects originate from backscattering-induced coupling between the CW and CCW mode families, which modifies the intracavity field dynamics and allows simultaneous comb generation in both directions.

Interestingly, when the CW–CCW interaction is sufficiently weak, it becomes possible to generate counter-propagating solitons by pumping the two directions independently \cite{yang2017Comb}. Although these solitons belong to the same spatial mode family, their round-trip times and hence their repetition rates can be controlled independently through the two pumps. Moreover, a regime exists in which both the relative optical phase and the relative repetition rates of the two soliton streams become mutually locked, see e.g. \cite{zhang2022bark-bright}. This phase- and repetition-rate-locked state enables a single microresonator to generate two coherent soliton frequency-comb streams with distinct repetition rates, providing a powerful platform for dual-comb spectroscopy and precision metrology. This configuration was studied in more detail in MgF$_2$ WGM resonators \cite{qureshi2023DualComb}.

%\paragraph{GVD optimization of group velocity dispersion in coupled resonators}
GVD of a microcavity system can be optimized by means of coupling of microresonators. It was shown theoretically that a photonic compound ring resonator can have a large anomalous GVD at any desirable wavelength, even if each individual resonator is characterized with normal GVD \cite{soltani16lpr}. Advancing the idea further it was found that making sets of two and three coupled microring resonators one can %significantly modify properties of the individual rings 
enable new properties of the generated frequency combs and create phase-locked solitons arrays \cite{yuan2023_3rings,Ji2025np}.

%\paragraph{Improved power efficiency}
As shown by \cite{bao14ol}, the comb efficiency generally degrades with the number of comb lines.  The broad Kerr comb harmonics %leaving the resonator 
have low contrast compared to the pump. Optimizing the resonator loading one is able to generate high contrast solitons of limited bandwidth \cite{Grudinin2017Optica}. Soliton generation by pulse \cite{Li:22} or dichromatic \cite{zhang2020dblcmb,Matsko2025ol} pumping provides a way to increase the conversion efficiency. Other efficiency improving techniques such as pump recycling in coupled resonators \cite{Xue2019NatPhotonics}, interferometric back-coupling \cite{ChavezBoggio2022NatCommun}, mode-shifting in photonic molecules \cite{Helgason2023NatPhotonics}, as well as photonic crystal reflectors \cite{Zang2025NatPhotonics} were introduced. The efficiency can be improved rather significantly by taking advantage of mode interactions between spatial modes within a single microresonator \cite{Liu2025arxiv2512.05090} as well as by involving a pair of optimally engineering microresonators \cite{Zhu2025arxiv2503.02022}. The last approach allows generation of a coherent octave spanning comb with a reasonable pump power and repetition rate, solving one of the outstanding problems of the technology.

%\paragraph{Kerr combs at shorter wavelengths}
Extending of the comb range into visible and ultra violet parts of the optical spectrum is challenging because of the normal GVD of the majority of optical materials in that frequency region as well as due to increase of the material attenuation at shorter wavelength. In the micro-fabricated systems scattering due to imperfections of the manufacturing process is particularly significant. A solution was suggested for a coupled microresonator system \cite{soltani16lpr,Dorche2017OptExpress,Liu2025NatCommun4780}. However, this configuration usually does not allow for generation of broadband frequency combs. A near-visible comb was observed in a pulse-pumped aluminium nitride microring resonator \cite{Guo2018PhysRevApplied}. The GVD improving technique based on the resonator morphology optimization was demonstrated \cite{Lee2017NatCommun,Ma2019PhotonicsRes,Li2025AdvPhotonics}. To our knowledge, neither of these techniques has led to generation of the combs in the UV frequency region. 

%\paragraph{Integration of the Kerr combs}
High power-efficiency of Kerr frequency comb generation is critical for their integration. This problem was solved in silicon nitride microresonators with moderate FSR \cite{xuan2016Si3N4}.
Integration of Kerr frequency combs into small form-factors without usage of bulky optical amplifiers was significantly improved recently. Packaged combs based on silicon nitride cavities were reported in \cite{Stern2018Nature,raja19nc,Shen2020Nature,Rizzo2023NatPhotonics,Ji2025np}. Ultimately these packages support battery-powered operation of the frequency combs \cite{Stern2018Nature} unless thermal stabilization of the microresonator is required.

Significant progress has been made in the on-chip integration of crystalline microresonator frequency combs over the past decade. For instance, these combs have been utilized in microwave photonic oscillators, where a self-injection-locked laser coupled to a MgF\textsubscript{2} microresonator produced signals with phase noise as low as –135~dBc/Hz at a 10~kHz offset from an 11~GHz carrier \cite{matsko16ifcs}. The same noise level was reached at the same offset frequency with a 20~GHz carrier by \cite{kudelin2024Photonic}. This chip-integrated source featured an optical comb stabilized with two external frequency references. 

A compact, environmentally robust Ka-band oscillators, implemented in 0.25~cm$^3$ low-temperature co-fired ceramic packages, were developed and tested, achieving –120~dBc/Hz phase noise at 10~kHz offset for carrier frequencies of 28~GHz and 35~GHz \cite{seidel18mwp}. To our knowledge, these results represent the highest spectral purity and the most mature technology demonstrated for photonic oscillators of sub-cubic-centimeter volume. More recently, this kind of platform has been extended to W-band signal generation \cite{savchenkov24LO}, achieving the same phase-noise performance for the 92.4~GHz carrier.

%\paragraph{Coherence properties of Kerr combs}
The coherence of individual harmonics in microresonator based frequency combs has emerged as a critical metric for precision photonics applications. While microcombs generate densely spaced optical lines across broad spectra, the phase stability of each harmonic determines the comb’s utility in  coherent communications \cite{marin-palomo2017comm}, spectroscopy \cite{zhang2025combs_rev}, ranging \cite{trocha2018Lidar,suh2018Lidar,riemensber2020comb}, optical waveform synthesizers \cite{spencer2018synth}, and optical clocks  \cite{newman2019clock}. The WDM communication approach by \cite{marin-palomo2017comm} has led to the aggregate rate of 37.2 Tbit/s over a 75 km fiber link using 93 channels across the L and C optical communication bands. Both the synthesizer and the clock \cite{spencer2018synth,newman2019clock}, as well as the LIDAR by \cite{trocha2018Lidar}, were based on a pair of soliton-generating resonators with different pulse rate, which allowed for an effective increase of the spectral frequency interval of the octave-spanning comb. The frequency doubling was used to enable the $f$--2$f$ phase locking, as well as the frequency locking to a rubidium vapour reference. The LIDAR by \cite{suh2018Lidar} used only a single ring resonator, with the two solitonic combs propagating in the opposite directions. The small, of the order of 6 kHz, difference in the clockwise and counter-clockwise soliton repetition rates emerged due to the nonlinear dependence of the group velocity on the pump detuning from the resonance.

Sufficiently broad frequency combs enable self-locking and complete frequency stabilization leading to optical clock development. The 2f-3f self-locking was demonstrated by \cite{brasch2017Selfreferenced} in a silicon nitride  on-chip micro ring resonator with optimal GVD. An octave-spanning comb in the soliton regime was demonstrated by \cite{li2017sol}, who saw single as well as multiple solitons. Generation of these solitons was helped by the dispersive Cherenkov waves, and required a precise GVD engineering. An even more versatile GVD control has been achieved by \cite{yang2016GVDengineer} in lithographically fabricated wedge resonators with rather extreme rim shapes.

\begin{figure}[htbp]
\centering\includegraphics[width=8.5cm]{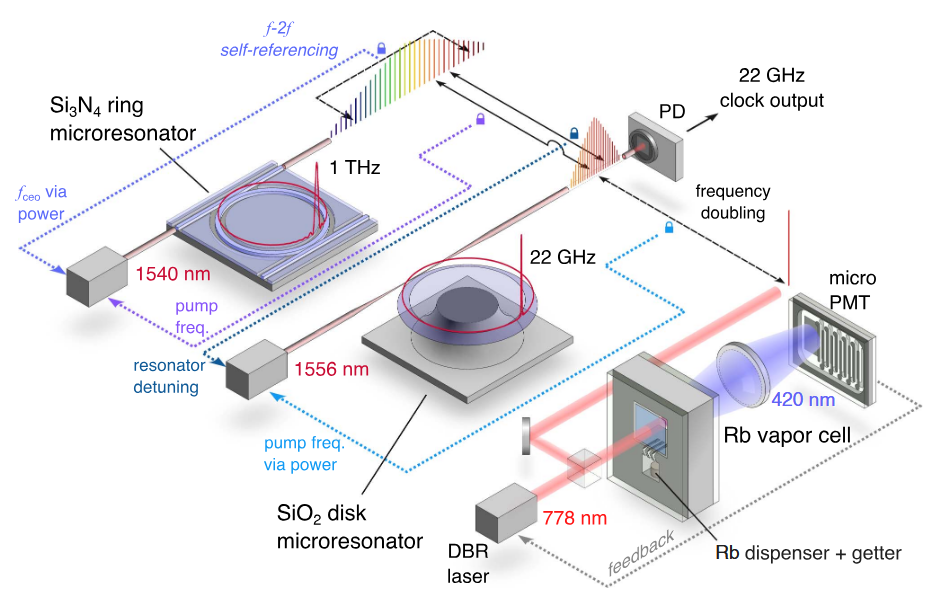}
\caption{An optical comb clock diagram using two micro-resonators with different soliton repetition rates, and an atomic frequency reference. Reproduced from \cite{newman2019clock}.
% This article publishe dunder the OSA Open Access Publishing Agreement allowing for non-commercial content reuse.
}
\label{fig:newman2019clock}
\end{figure}

The optical clock diagram by \cite{newman2019clock} is shown in Fig.~\ref{fig:newman2019clock}.  Here, a silicon nitride chip-integrated dispersion-engineered resonator with $R=23\,\mu$m generated an octave-spanning, single-soliton $f$--2$f$ locked comb. This clock achieved the fractional stability of $4.4\times 10^{-12}/\sqrt{t}$, with the floor of $10^{-13}$. This performance falls quite short of the large optical comb clocks, but the small size and low power consumption creates a special niche for the micro-resonator comb clocks.

Early studies showed that self-injection locking of the pump laser to high-Q crystalline microresonators could substantially reduce phase noise and synchronize individual harmonics across the comb \cite{arafin2017optiexp,arafin2017ieeepj}. Soliton formation provides a natural mechanism for locking the phases of all comb lines, producing harmonics with coherence approaching that of the pump laser itself. The unavoidable noise associated with the comb generation contaminates spectral purity of the harmonics \cite{lei2022natcommun3161}, but it can be negligible in combs based on crystalline microresonators. Interestingly, the frequency comb formation mechanism also can lead to the reduction of the technical noise of the individual optical harmonics below the noise of the original pump light \cite{lei2022natcommun3161}.

%\paragraph{Kerr combs with dual pump}
Dichromatic coherent excitation of a nonlinear optical cavity gives rise to a wide spectrum of nonlinear effects that do not occur in the conventional single-pump configuration. Among the examples are realization of dissipative optical time crystals \cite{taheri22nc}, Kerr nonlinearity-mediated synchronization of dissipative Kerr solitons \cite{moille25o}, and trapping of dissipative solitons \cite{taheri17epjd,taheri25lpr}. These results demonstrate how multiple driving fields fundamentally enrich the nonlinear dynamics of optical resonators, creating new directions for research in nonlinear and quantum photonics with cavities driven by polychromatic sources.

The four-wave mixing (FWM) involving a single signal and idler harmonics has been observed with a monochromatic pump in a photonic crystal resonator  \cite{marty21np} and in a multimode microresonator operating in the vicinity of the oscillation threshold \cite{dutt15pra}. A related configuration based on the GVD management of a ring resonator was suggested in \cite{matsko16ol}. Spatial separation of the co-propagating harmonics emitted by the resonators typically relies on optical filters, which inevitably introduce insertion loss undesirable  for quantum light generation. A counter propagating geometry based on dichromatic pumps can overcome this limitation by separating the two outputs in space.

In most cases where two counter-propagating pump waves are resonant with cavity modes, the FWM process does not satisfy phase matching conditions. These configurations have been exploited for other purposes, e.g., thermal stabilization of the resonator generating a Kerr frequency comb without altering the Kerr comb dynamics \cite{zhou19lsa}. The dichromatic pump can also induce spontaneous symmetry breaking in the nonlinear cavity, but in ways that are not directly linked to the FWM process \cite{del_bino2017PT}. 

\subsection{Third harmonic and 3PDC }

Third harmonic generation (THG) is a highly-nonlinear process whose power efficiency scales approximately as the fourth power of the resonator $Q$-factor and inverse square of the mode volume \cite{surya2018THG}. This gives small high-$Q$ resonators a tremendous advantage. The disadvantages are a poor overlap between the fundamental and third-harmonic eigenfunction, and the difficulty in attaining the required phase matching. While in the SHG the chromatic dispersion can be compensated by the polarization and/or geometric dispersion (in small resonators), the chromatic dispersion between the fundamental and third-harmonic wavelengths is usually too large. One solution to this problem was developed by \cite{surya2018THG} who fabricated a composite silicon nitride -- aluminum nitride microring resonator whose effective chromatic dispersion could be compensated for relatively low-order WGMs: the fundamental (1542 nm) TE mode with the radial order $q=1$ and a third-harmonic TE mode with $q=3$ (i.e., two nodes in the radial direction). With this resonator, the authors achieved a THG efficiency of 180\%/W$2$, corresponding to an absolute conversion efficiency of 0.16\%.

Alternatively, the WGM resonator dispersion can be modified via the cross-phase modulation in a Kerr medium. This technique was used by \cite{fujii2017THG} to facilitate the THG phase matching concurrent with an optical comb generation. In this case one of the comb lines played the role of the THG pump.

THG from a 1553 nm pump was also observed in a microbubble resonator~\cite{liu_monitoring_2025}, reaching the conversion efficiency of $2.6\times 10^{-9}$ at 1.5 mW pump power. In this case the phase-matching was achieved by leveraging the thermo-optical effects induced by the pump. The thermo-optical phenomena is in many respect similar to the Kerr phenomena (on a different time scale), so one may draw an analogy between the approaches by \cite{liu_monitoring_2025} and \cite{fujii2017THG}. Let us also mention an extensive review by \cite{jiang2020Optothermal} dedicated to the thermo-optical phenomena in WGM resonators.

Closely related to THG is the third-order sum-frequency generation, which can be achieved either through the cubic nonlinearity or the cascaded quadratic nonlinerity. The former type of optical frequency up-conversion was demonstrated in silicon nitride microrings by \cite{lu2021freq_conversion}, the latter, in the x-cut lithium niobate microrings by \cite{lin2019Broadband}.

Degenerate third-order parametric down conversion, the 3PDC, is the process opposite to the THG, when a short-wavelength photon breaks down into a triplet of identical longer-wave photons. The photon energy and momentum (orbital momentum in WGM resonator geometry) must be conserved in this process. The 3PDC can also be non-degenerate, when the generated photons have different frequencies.

The interest in this process was inspired by the unique quantum states it may access. The progress in this field remains largely theoretical, see e.g. \cite{okoth20193PDC}. To-date, 3PDC in optical resonators remains elusive, although it has been observed in microwave SQUID resonators \cite{chang2020RF3PDC}.

\subsection{Raman and Brillouin scattering}
Inelastic light–matter interactions such as Raman and Brillouin scattering provide complementary pathways for frequency conversion in WGM resonators. Unlike parametric processes discussed above, these interactions rely on coupling to internal material excitations (optical phonons in the case of Raman scattering and acoustic phonons for Brillouin scattering) leading to distinct coherence properties, gain spectra, and dynamical behaviour. Farnesi et al discuss ``phoxonic'' cavities that support photons (optical) and phonons (acoustic) in a review paper focusing on hollow ~\cite{farnesi_efficient_2017} and glass~\cite{farnesi_phoxonic_2021} WGM cavities.

\paragraph{Raman}
There have been a number of works studied the competing effects of Kerr and Raman on the formation of combs in WGM cavities. Fujii et al.~\cite{fujii_transition_2018} experimentally demonstrated controlled shifting from the FWM modulational instability domain to Raman scattering to generate a broad stimulated Raman scattering (SRS) comb in a silica WGM cavity. The role of Raman scattering on resonator soliton dynamics was theoretically explored by \cite{wang2018soliton}, where the authors show that if the Raman linewidth is broader than the soliton bandwidth, the Raman effect can limit the comb spectral width and consequently, the soliton minimal duration. However, when the Raman linewidth is narrower than the comb spacing, Raman lasing can compete with Kerr comb and disrupt comb formation \cite{okawachi_competition_2017} and cavity-soliton formation. A Stokes soliton can coexist with the fundamental dissipative Kerr soliton if they are based on different mode families, as observed by \cite{yang_stokes_2017} in a silica wedge microresonator. 

Jung et al~\cite{jung_stokes_2019} also demonstrate that a formed Kerr comb can itself act as a multi-frequency pump for Raman processes in an AlN microring. A broadband (wider than \SI{242}{nm}) two-dimensional (2D) Kerr and Raman–Kerr frequency comb was shown in \cite{jin_controllable_2021} with reversible switching between a FWM state and a SRS state and an octave spanning Raman-assisted frequency comb was reported in \cite{weng_octave-spanning_2021}. Last year, the controlled conversion from a pure Raman frequency comb to a Raman-assisted frequency comb via coupling tuning resulting in dispersion modification in a packaged silica microbottle resonator was also demonstrated in \cite{wang_numerical_2025}.

A mid-IR Raman comb (up to \SI{2.7}{\micro m}) was shown in a silica bottle resonator that served as a mode reflector for the thulium-doped fiber lasing as well as a nonlinear initiator for the production of Raman lasing~\cite{jiang_mid-infrared_2020}. A scanning microprobe using WGM‑enhanced Raman spectroscopy was implemented in a configuration similar to an atomic-force-microscopy (AFM) tip~\cite{mao_whispering-gallery_2023}. They demonstrate an improvement of two orders of magnitude in sensitivity compared to purely plasmonic sensors. \cite{fan_high-specificity_2023} demonstrate WGM sensing that identifies the chemical being detected by measuring the Raman signal, i.e., the ``fingerprint spectrum'' of surface-enhanced Raman scattering, rather than simply tracking a change in WGM resonance parameters.

In \cite{kovach2020Raman}, the authors modify the surface of a silica resonator with an organic monolayer to increase the Raman gain and observe four orders of stimulated Raman peaks with sub-mW input power. Furthermore, the stimulated anti-Stokes Raman scattering efficiency is improved by three orders of magnitude compared to previous studies with hybrid resonators. Cascaded continuous-wave Raman lasing extending to \SI{2}{\micro m} via a telecom pump was shown in an AlN-on-sapphire microring integrated platform~\cite{liu_integrated_2017}.

\paragraph{Brillouin} Brillouin lasers typically have narrow linewidths due to their narrow gain profile. In WGM microcavities, this can be further narrowed (\SI{100}{Hz}) in exchange for robust all-optical locking as reported in~\cite{yang_bridging_2018,mao_narrow-linewidth_2025}. Cascaded Brillouin lasing, on the other hand, has been shown to broaden the fundamental linewidth of the first-order Brillouin laser~\cite{wang_cascading-induced_2025}. Theoretical investigations into the dependence of acoustic wave velocity (and therefore, the Brillouin gain of the medium) on the crystalline axes’ orientation~\cite{tian_theoretical_2024} indicated that z-cut MgF\textsubscript{2} WGM resonators could feature the widest Brillouin gain range of approximately 2.6 GHz. This was experimentally validated in a demonstration of stimulated Brillouin scattering (SBS) with Brillouin shifts of \SIrange[]{11.73}{14.47}{GHz}~\cite{lin_hundredfold_2023}. Furthermore, that device was shown to support a Kerr-Brillouin comb spanning \SI{120}{nm}. Cascaded Brillouin lasing and Brillouin-Kerr combs spanning \SI{250}{nm} was also demonstrated in a similar MgF\textsubscript{2} WGM resonator~\cite{xu_cascaded_2023}. This process was also shown to lead to a 107 GHz soliton formation in silica microresonator \cite{zhang2025BrilComb}.

MgF\textsubscript{2} resonators are of high interest as platforms for lasing based on both SRS and SBS~\cite{tian_magnesium_2024} due to the high-Q achievable in them (exceeding \SI{1e8}{} at telecom wavelengths) as well as their transparency extending into the mid-IR. A milestone result showed sub-Hz linewidth Brillouin lasing on-chip~\cite{gundavarapu_sub-hertz_2019}. Other reports on Brillouin lasing include heterogenously integrated chip devices with on-chip slope efficiency of 5\%~\cite{yu_brillouin_2024} and lasing in the mid-IR domain~\cite{pathak_microresonator_2024,ko2024Brillouin}. A microcavity Brillouin laser was also used as a sensor for external light, acoustic and microwave sources~\cite{yang_multiphysical_2020}.

SBS has also been used to optomechanically cool optical modes (conversely, amplify specific mechanical modes in the medium) as discussed in~\cite{shen_dissipatively_2021}. There have also more unique explorations involving Brillouin scattering in dielectric resonator platforms such as the implementation of a parity-time phase transition near an exceptional point using forward SBS~\cite{chen_exceptional_2022}. Furthermore, the constraint of resonator size due to phase-matching requirements for SBS can be lifted by the inscription of a Bragg grating as shown in~\cite{wang_taming_2024}, where phonon lasing was demonstrated in a cavity of size \SI{20}{\micro m}.

\subsection{Optomechanics}

As we discussed above, Raman and Brillouin scattering can facilitate a strong coupling of light with the optical or acoustic phonons. The quantum regime of such a combined optomechanical system can be very interesting. Indeed, there has been a proposal \cite{Peano15optomech} to couple mechanical oscillations in a WGM resonator to the squeezed light internally generated in this resonator. However due to very low phonon energies and their strong coupling to the thermostat, the thermal phonon population is always present and is masking any quantum phenomena. To avoid this, one has to cool the resonator to an extremely low temperature, ``freezing out" the thermal phonon population and reaching the phonon mode's ground state. This approach, involving a dilution refrigerator cooling a macroscopic quartz crystal to 0.2 Kelvin, was taken by \cite{doeleman_brillouin_2023}. The thermal phonon population below 0.5 quanta per mode was thereby achieved. 

\begin{figure}[htbp]
\centering\includegraphics[width=8.5cm]{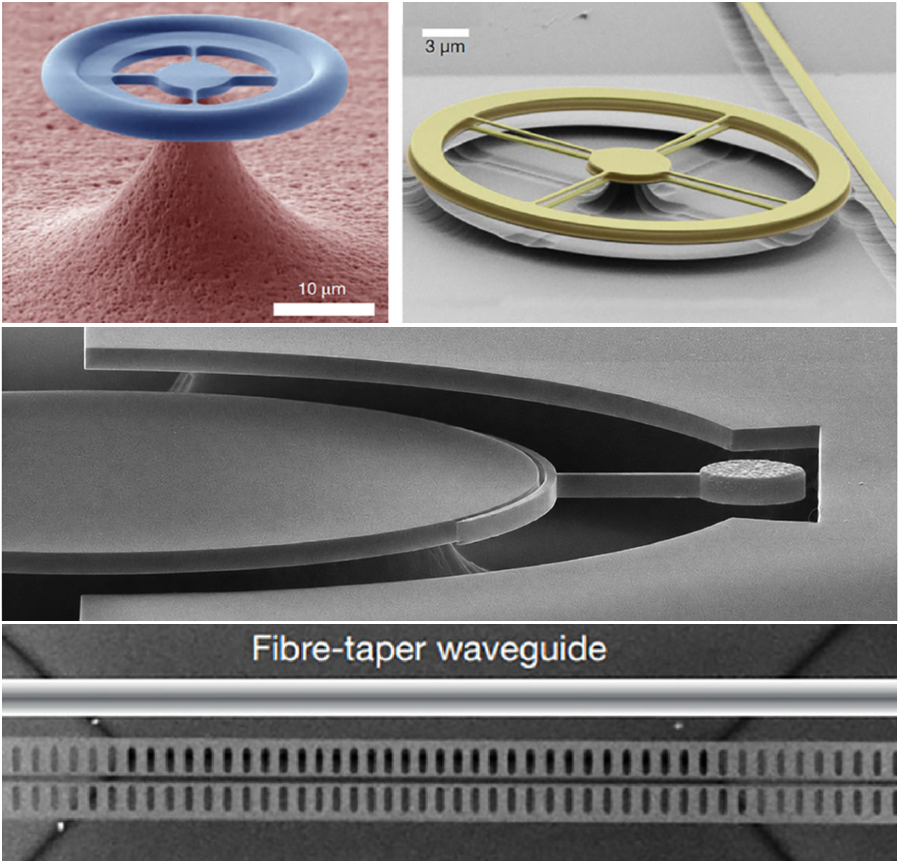}
\caption{Representative images of various types of optomechanical oscillators thermally and mechanically decoupled from the environment. Reproduced from \cite{li2021rev} with modifications.
%This article is licensed under a Creative Commons Attribution 4.0 International License, which permits use, sharing, adaptation, distribution and reproduction in any medium or format
}
\label{fig:Li2021revFig3}
\end{figure}

In an alternative approach, \cite{dania2025optomech} decoupled a silica nanoparticle from the room-temperature lab environment by optically levitating it in vacuum. The particle was then optically cooled to approximately 0.04 thermal quanta. This cooling was associated with a librational oscillation of a single particle rather than a phonon mode, but this demonstration highlighted the advantages of the environment-decoupling approach. Such a decoupling can be achieved in artificially created micro-structures, such as shown in Fig.~\ref{fig:Li2021revFig3} designed to function as both a mechanical and an optical resonator at the same time. The strong decoupling of the mechanical modes from the thermostat, and their high $Q$-factor, allow for efficient optical cooling. 

Modern optomechanics is a large and rapidly growing field. Its progress since 2016 is discussed in review papers such as \cite{li2021rev,barzanjeh2022optomech,wen2023cooling,croquette2023rev}. We will not attempt to present this vast amount of research here, but will briefly mention some of the early forays into the domain of quantum optomechanics.

Two-mode quantum correlation between optical and mechanical oscillators quadratures was studied by \cite{shomroni2019backaction}, who detected 0.67 dB of squeezing performing the \textit{backaction-evading} measurement. The high-$Q$ acoustic modes have been considered as intermediary for the microwave-to-optics coherent conversion and quantum state transfer by \cite{higginbotham2018conv}. Using a membrane coupling an optical resonator with a microwave LC circuit, they demonstrated the coherent conversion of 6 GHz microwaves to 282 THz light with the photon-number efficiency of 47\% and approached the quantum regime. This technique however required a deep cryogenic environment of 35 mK.

\section{Lasers}\label{sec:laser}

Using narrow-line, tunable, mechanically stable whispering gallery resonators for laser stabilization is perhaps their most obvious application. Indeed, the WGM lasers date back to 1960s \cite{garrett_stimulated_1961,Walsh63laser} and continue to make progress. This progress is stimulated by many important applications, including those reviewed by \cite{reynolds2017fluorescent,li2026WGMlasREV}.  

As the most direct approach, a passive resonator mode can be used as a reference, and an external laser can be frequency-locked to it using an electronic locking method such as e.g. the Pound-Drever-Hall technique. This approach has been actively explored, allowing for reaching the fundamental noise limit with the temperature-compensated ``sandwich" resonator we mentioned earlier \cite{lim2017sandwich}. This limit depends on the resonator parameters, and in this case corresponds to approximately 25 Hz laser linewidth.

Other approaches involve the optical injection locking of a laser to the reference resonator, or integration of the laser gain media with the resonator itself. 

%\subsection{\color{red}Locking external laser to a passive resonator}

\subsection{Self-injection-locked semiconductor lasers}

The rapid maturation of precision photonics has brought laser phase and frequency fluctuations into the foreground of both metrology and information technology. In particular, optical atomic clocks operating at the forefront of time and frequency standards \cite{chabuda16njp,leroux17m} and modern coherent fiber-optic links designed for extreme data throughput \cite{kakkar17sr} have made it impossible to ignore the detailed spectral structure of laser noise. These applications are no longer limited by simple white or Lorentzian linewidth models; instead, performance is governed by how phase noise is distributed over Fourier frequency, motivating a more nuanced theoretical and experimental treatment \cite{tran19aplp}. In coherent communication systems, it has been shown that channel capacity is constrained by the power-law behaviour of laser phase noise, typically following an $f^{-\alpha}$ dependence with $\alpha>2$ \cite{kakkar17sr}. Likewise, in optical clocks the long-term stability of the clock laser, especially when locked to an ultra narrow atomic transition, is ultimately restricted by flicker-type frequency noise, which directly degrades both timekeeping stability and the fidelity of clock-to-clock comparisons \cite{chabuda16njp,leroux17m}.

\begin{figure}[htbp]
\centering\includegraphics[width=8.5cm]{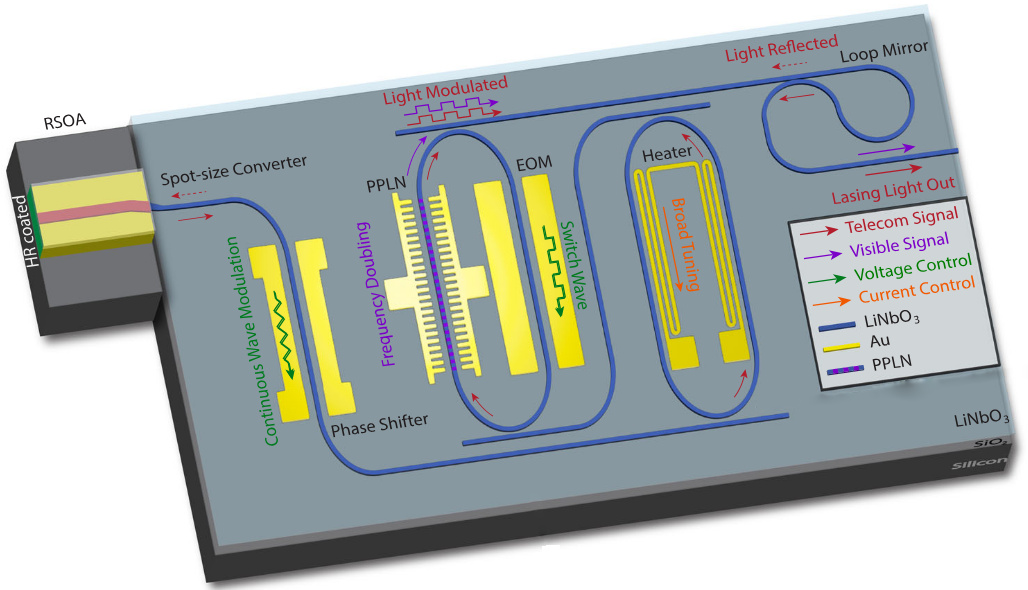}
\caption{A PIC schematic combining various photonic elements on a single chip, used by \cite{li2022laser} for a self-injection-locked laser. Reproduced from \cite{li2022laser} with modifications.
%This article is licensed under a Creative Commons Attribution 4.0 International License, which permits use, sharing, adaptation, distribution and reproduction in any medium or format
}
\label{fig:li2022laserFig1a}
\end{figure}

Parallel to these application-driven demands, significant progress has been made in compact low-noise laser sources. One particularly successful strategy has been the migration of microresonator-based self-injection-locked (SIL) semiconductor lasers into photonic integrated circuit (PIC) platforms \cite{jin2021inj_lock,xiang21arch,Stern2018Nature,li18jlt,raja19nc,raja20oe,li2022laser,snigirev2023laser,white2024laser}, such as illustrated in Fig.~\ref{fig:li2022laserFig1a}. Originally developed using crystalline whispering-gallery-mode resonators (WGMRs) \cite{liu2018waveguide}, this architecture has demonstrated record-low flicker noise for miniaturized external-cavity semiconductor lasers. 

At its core, self-injection locking enhances the coherence of a laser without relying on high-bandwidth electronic feedback or bulky external cavities. % \cite{hjelme91jqe,dahmani87ol,hollberg88apl,li88apl,li89jqe,li89jqe1,laurent89jqe,himmerich94ao,vassiliev98oc,vassiliev03apb,liang10ol,ilchenko11spie,burkart13ol,liang15ao,liang15nc,savchenkov2018selflock,donvalkar18jo,pavlov18np,galiev18oe,jin2021inj_lock,yacoby2021injlock}. 
Instead, a passive, frequency-selective optical cavity reflects a small fraction of the emitted light back into the laser chip, where it interferes with the intracavity field and enforces better phase coherence. When monolithic microresonators are used, this feedback is most often generated by resonant Rayleigh back scattering intrinsic to solid-state materials and enhanced due to the resonant interaction \cite{Weiss1995splitting,gorodetsky2000}. Recent examples of such lasers include
\cite{savchenkov2018selflock,donvalkar18jo,pavlov18np,galiev18oe,jin2021inj_lock,yacoby2021injlock}. Also see a comprehensive review by \cite{kondratiev2023rev}.

Crystalline toroidal WGMRs remain particularly appealing for SIL because they offer exceptionally high $Q$ factors, low thermorefractive noise, and relatively large mode volumes. Their main drawback is the weakness of intrinsic Rayleigh backscattering, which can limit the achievable locking range. This issue has been addressed through controlled surface roughening or by introducing dielectric or metallic nanoparticles to enhance scattering \cite{savchenkov20sr}. 

SIL lasers have proven valuable across a wide range of photonic technologies. They provide compact sources of highly coherent, tunable light for spectroscopy and ranging, generate spectrally pure microwave and millimeter-wave signals for radar and communication systems, and significantly improve the stability of optical sensors. In addition, they serve as low-noise pumps for Kerr-based frequency-comb generators \cite{savchenkov16np,gaeta2019CombRev,kovach20aop,savchenkov24LO}.

From a theoretical standpoint, early models of SIL \cite{Kondratiev:17} captured only a limited subset of the relevant physics. A comprehensive treatment was later developed in \cite{galiev20pra}, which identified five key parameters governing locking performance: (i) the strength of counter-propagating mode coupling set by resonator backscattering; (ii) the locking phase determined by the laser–resonator optical path and the resonator eigenfrequency; (iii) the absolute optical path length between the laser and the resonator; (iv) the frequency detuning between the free-running laser and the cavity mode; and (v) the pump–resonator coupling efficiency dictated by spatial mode matching. Systematic exploration of this multidimensional parameter space revealed a global optimum and, importantly, showed that increasing the backscattering beyond a certain level yields no further improvement in stabilization—an insight critical for practical device design.

The theoretical treatments clarified that the performance of SIL depends critically on the quality factor of the resonator providing feedback. They also highlighted the need for low modal density and exceptional mechanical stability of both the resonator and the optical path connecting it to the laser. Because conventional mirror-based cavities are sensitive to vibration and thermal drift, their practical usefulness for SIL has remained limited.

\subsection{Lasing in optically active resonators}\label{sec:oalas}

Lasing in optically active microring and whispering gallery mode resonators provides an attractive approach to generating coherent radiation by combining optical gain with strong resonant confinement and frequency-selective optical feedback. The use of high-$Q$ dielectric resonators ensures long photon lifetimes and control over the spectral and modal properties of laser emission. %Rare-earth ions, semiconductor quantum dots, color centers, and other active materials can be incorporated into dielectric resonators, extending the range of accessible wavelengths and gain mechanisms beyond those readily available in conventional semiconductor lasers. 
Whispering gallery lasers (WGLs) are attractive for such applications as precision metrology, optical frequency references, and microwave photonics. They also have outstanding mode locked properties and offer a promising route toward multifunctional coherent photonic sources that combine laser operation with frequency stabilization and nonlinear signal processing. Resonator arrays and coupled-resonator architectures offer additional opportunities for engineering the spatial and spectral properties of the emitted light.

Fabricating the WGLs from optical gain materials continues to be an active research area. As one approach, rhodamine 6G, typically pumped around 520 nm and lasing in the range of 560 - 610 nm, was dissolved in clear polymers and turned into droplet micro-resonators suspended on an optical fiber like dew on spider web \cite{Xie2017laser,gu2017laser}. The single- and multi-mode lasing was observed when these resonators were flood-pumped by 10 ns laser pulses.

Another class of broadly used gain materials is the perovskites. These materials have a general chemical formula ABX$_3$, where A and B are cations and X is an anion, usually oxygen or halide. The microcavities with $Q\approx 5000$ made from CsPbX$_3$ (where X represented various combinations of chloride, bromide and fluoride) were used by \cite{zhang2016high} to observe lasing in the range of 400 to 700 nm. These microcavities have a rectangular shape and cannot be properly described as whispering galleries, but they do support high-$Q$ modes confined by the total internal reflection. Low-threshold lasing at 556.4 nm was demonstrated in  a CH$_3$NH$_3$PbBr$_3$ perovskite micro-rod resonator by \cite{wang2017whispering}. This resonator also had $Q\approx 5000$. It was pumped with a frequency-doubled Ti:sapphire laser at 400 nm. A different approach was taken by \cite{price_whispering-gallery_2020} who coated silica micro-spheres with CsPbBr$_3$ and CsPbI$_3$ nanocrystals. This WGL was also flood-pumped by optical pulses, lasing at approximately 520 nm and 700 nm for CsPbBr$_3$ and CsPbI$_3$, respectively.

The third group of WGL are those based on rare-earth-doped optical materials. The theoretical study of such lasers is presented by \cite{ceppe2019laser}. Experimentally, continuous-wave single-frequency lasing in an LED-pumped neodymium-doped YVO$_4$ disc resonator was demonstrated by \cite{Herr2017laser}. The lasing wavelength was 1064 nm and power 0.8 mW with a 3.5 W pump flooding the resonator. Later this system was modified to allow for the lasing frequency tuning with a mode-hop-free range of 11 GHz \cite{herr2019tunable}. A single-frequency laser based on a neodymium-doped \LN{} microresonator was demonstrated by \cite{minet2022EOlaser}. This laser was pumped at 813 nm and emitted light at approximately 1086 nm, with the pump threshold of 0.3 mW.  Notably, it was pumped by exciting a resonator's WGM instead of flooding it with external light. The laser was electro-optically tunable in the range of 3.5 GHz. The same resonant-pumping approach was used with an erbium-doped \LN{} WGL by\cite{wang2021laser,yin2021laser}, leading to a similar performance. In a later implementation by \cite{li2023optically}, a tiny (30 $\mu$m diameter) WGL reached the slope efficiency of 12.4\% with a maximum output power of \SI{1.12}{\milli\watt} in the multi-mode lasing regime, and the slope efficiency of 6.18\% with a maximum power of \SI{0.5}{\milli\watt} in the single-mode regime. Notably, the resonator in this case had a slightly eccentric shape, which allowed for the free-space out-coupling of the laser radiation.

Both single-mode and multi-mode lasing regimes were also observed in a titanium-doped sapphire WGL by \cite{azeem2022lasing}. Distinctly from the flood-pumping approach, this WGL was pumped with a 516.6 nm laser coupled into a single WGM. It had an exceptionally low threshold of 14.2 mW, with the slope efficiency of 34\%. This resonator was furthermore shown to operate as an amplifier for a 795 nm signal.

\begin{figure}[htbp]
\centering\includegraphics[width=8.5cm]{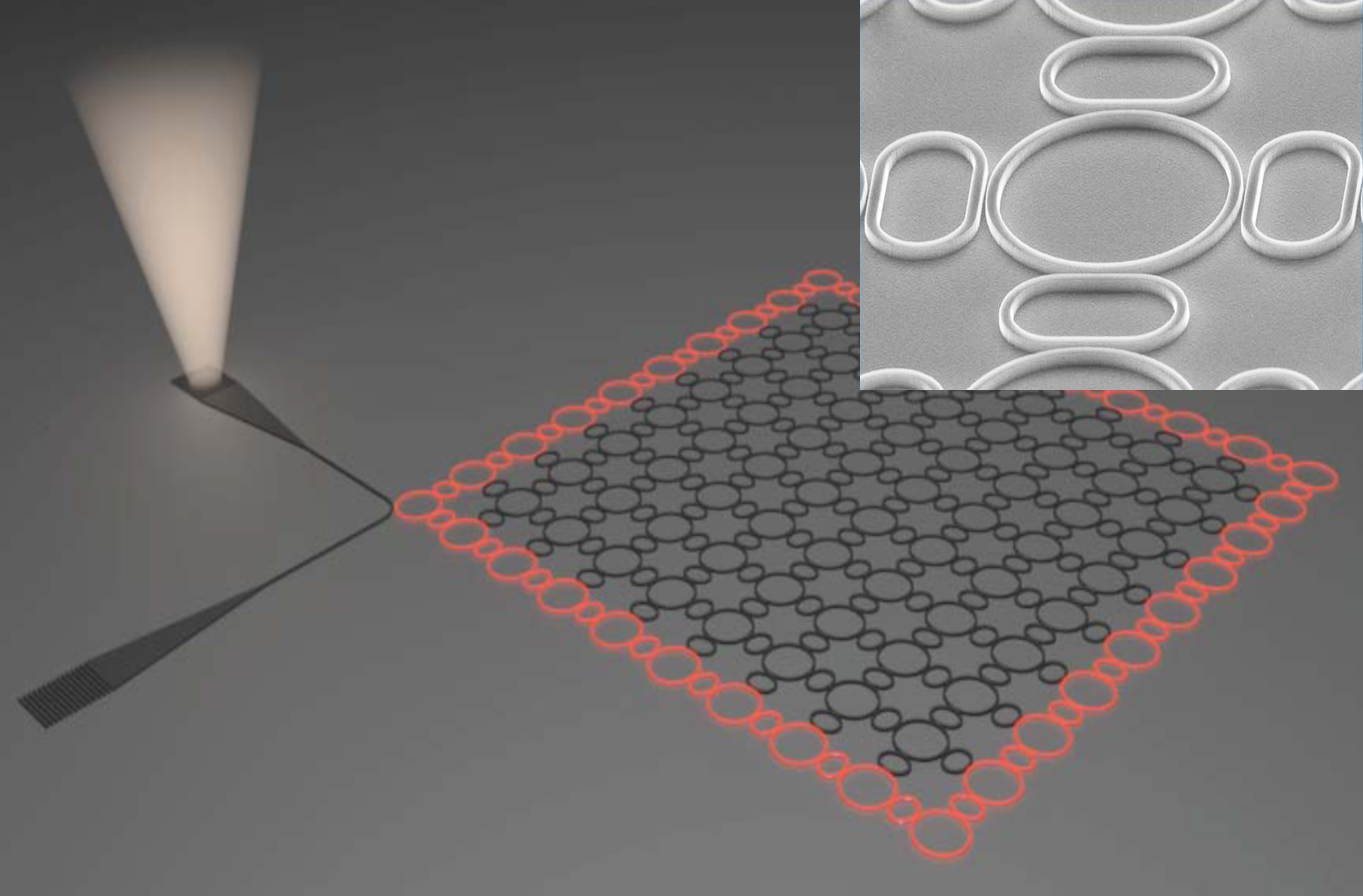}
\caption{A topological insulator laser schematic with an SEM image of the unit cell. The topologically protected ``meta-mode" locus is highlighted in red. Reproduced from \cite{bandres_topological_2018} with modifications.
% Sharing license on file
}
\label{fig:bandres_topological_2018_Fig1}
\end{figure}

Resonator \textit{arrays} allow for implementing of WGLs with unique properties. A topological insulator laser was theoretically discussed \cite{Harari2018topological} and demonstrated \cite{bandres_topological_2018} in a form of a 10x10 square array of racetrack and ring resonators supporting a single unidirectional laser ``meta-mode", topologically protected against individual resonators' defects. This array is schematically represented in Fig.~\ref{fig:bandres_topological_2018_Fig1}, with the unit cell's SEM photo shown as the inset.

Pulsed Q-switched Cr,Nd:YAG WGLs were demonstrated by \cite{wang2025laser} in both a micro-ring and photonic molecule configurations, with the typical pulse rate and duration of 21 MHz and 5 ns, respectively. Here the $Q$-factor modulation arises from the saturable absorption property of the Cr$^{4+}$ ions, while the Nd$^{3+}$ ions provided the optical gain. The pumping was again achieved by exciting a single WGM through a tapered fiber coupler, while the photonic molecule structure facilitated the vernier effect allowing for the single-mode lasing. This structure consisted of two side-coupled YAG WGM resonators of different diameters, one doped with Cr and Nd and performing the gain and saturable absorption functions, the other undoped and performing the function of the mode filter.

The presence of optical loss and gain in coupled WGM resonators can lead to the non-Hamiltonian dynamics and give rise to the exceptional points \cite{ozdemir2019PT}. For the WGLs, the exceptional point may serve as an effective saturable absorber and lead to a Q-switched pulsing laser, as proposed by \cite{imamura2024EP}. The PT-symmetry violation associated with the exceptional points can also result in spontaneously arising chirality \cite{cao2017chirality} and a controllable unidirectional lasing. Such a lasing was observed by \cite{peng2016Chiral} in a single resonator where the clockwise and counter-clockwise directions were associated with the gain and loss, respectively. It was shown to improve the sensitivity of the optical WGM-based nanoparticle detectors \cite{chen2017EP} and gyroscopes \cite{silver2021gyro}. Unidirectional lasing was also proposed by \cite{yang2020DP} in a photonic molecule, where one of the resonators was providing the gain, and the other, the loss.

Finally, deviating from the WGL topic, we would like to mention that the exceptional points and the PT-symmetry violation can be observed in completely passive systems, such that have only loss and no gain. Such a system would need a constant supply of energy. In photonics, this was realized by \cite{wang2020EP_EIT} who coupled a micro-disk and a micro-toroid resonators to a shared waveguide providing a constant optical pumping. In the PT-symmetric regime this system has four non-degenerate (due to the CW-CCW coupling) eigenfrequencies. However if for one of the resonators the PT symmetry is broken, one has a three-level system instead. Its level structure can support the photonic equivalent of the electromagnetically induced transparency (EIT) and absorption (EIA) regimes, associated with  the ``slow light" and ``fast light", respectively. 

\section{Quantum optics with resonators}\label{sec:quant}

\subsection{Optical squeezing}\label{sec:2sqz}

\paragraph{Single-mode squeezing}
The term \textit{quantum squeezing} applies to a dynamical system whose phase-space ``footprint" represented by a quasi-probability distribution is compressed in one direction beyond the classical limit. This direction is determined with respect to the canonical $(x,p)$ coordinates, or quadratures, corresponding to the quantum operators 
\begin{eqnarray}\label{quadr}
\hat{p}& =& -i \sqrt{\hbar \omega / 2} \left(\hat a  - \hat a^\dagger \right) \\
\hat{x} &= & \sqrt{\hbar / (2\omega)} \left(\hat a  + \hat a^\dagger \right)\nonumber
\end{eqnarray}
expressed through the photon creation and annihilation operators for this mode. The squeezing direction may coincide with one of the quadratures, or correspond to the angle or distance from the $(x,p)$ plane origin.

The main significance of the squeezed states comes from their ability to facilitate sub-shot-noise measurements. For example, using a squeezed vacuum state as a local oscillator one can perform an enhanced-sensitivity phase measurement. In our previous review \cite{Strekalov16rev} we listed some of the practical applications of this type of measurements, and discussed the key aspects of the single-mode squeezing in WGM resonators. Two of these aspects were highlighted as strong benefits of the WGM platform. One is that the spectrum of such squeezing may extend to very low frequencies, potentially enabling the sub-shot-noise interferometric measurements in the sub-kHz range. Such a low-frequency squeezing is very important for astronomy and astrophysics observations. The second aspect is a high purity of such squeezed states. This means that the squeezing along one phase-space direction and the inevitable anti-squeezing along the orthogonal direction closely preserve the minimal phase-space footprint area determined by the uncertainty relation $\Delta x\Delta p=\hbar/2$. In other words, the high-$Q$ WGM-based squeezers do not introduce a significant decoherence to the input optical state and preserve its quantum purity much better than the lower-efficient alternatives that require a strong optical pump. 

Decoherence of quantum-optical states is an important problem in quantum communication and data-processing protocols which reduces their fidelity. In addition to an imperfect source, it may arise from the photons loss and dephasing during propagation and detection. Random photon loss will destroy the photon-number squeezing, effectively superposing the initial squeezed state with the vacuum state as if with a beam splitter.  The phase noise which may be caused by thermodynamical and acoustical fluctuation of the index of refraction of the propagation media similarly affects the phase squeezing. These limitations and methods for their mitigation are discussed by e.g. \cite {serafini2005decoher, riccardi2021decoher, mele2024decoher}.

Further development of the WGM parametric squeezers was mainly built around these aspects. Squeezed vacuum demonstrated with a square-shaped monolithic resonator \cite{Brieussel16sqz} yielding a squeezing figure -2.6 dB at 5 MHz frequency offset, and with a WGM resonator \cite{otterpohl2019sqzd_vac} showing -1.4 dB at 0.5 MHz offset. These papers also discuss the limitations preventing the WGM approach from showing more spectacular squeezing figures.

Remarkably, \cite{otterpohl2019sqzd_vac} reports a substantial vacuum squeezing at the pump power exceeding the OPO threshold $P_{\rm th}=1.35\, \mu$W by a factor of approximately 1.6. The parametric light power emitted from the resonator at this point is still low enough to be measured with the same low-noise, high-efficiency photodiodes that are used for the sub-threshold measurement. The measurement is performed in a standard balanced heterodyne technique, when the signal is mixed with an optical local oscillator on a 50/50 beam splitter, and sent to two quadrature detectors. The squeezing and anti-squeezing are observed in the photocurrents difference as the local oscillator phase is varied. The observed above-threshold squeezing contradicts the common understanding that vacuum squeezing of a degenerate OPO should disappear above the threshold due to spontaneously arising state displacement \cite{Andersen16sqz_rev}.

\paragraph{Two-mode squeezing}
Two-mode squeezing is another name for continuous-variable entanglement between two modes. Instead of a squeezed quadrature of a single-mode quasi-probability distribution function, it is manifested by a stronger-than-classical correlation between two optical fields. Similarly to single-mode squeezing, the two-mode squeezing can be mapped on a phase-space plane if we redefine the single-mode operators in (\ref{quadr}) as combinations of two-mode operators:
\begin{equation}\label{quadr2}
\hat a=(\hat a_1+\hat a_2)/\sqrt{2},\quad\hat a^\dagger=(\hat a_1^\dagger+\hat a_2^\dagger)/\sqrt{2}.
\end{equation}

Like SHG and PDC, modern experimental research of two-mode squeezing largely migrated towards chip-integrated resonators. Two-mode squeezing was demonstrated in silicon nitride micro-rings \cite{zhao2020sqz,yang2021CombSqz,zhang2021Squeezed} by various research groups, all reporting similar squeezing figures ranging from -1.34 to -1.65 dB. 

The two-mode squeezing based on non-degenerate spontaneous PDC (SPDC) was taken to the extremely non-degenerate regime by \cite{sahu2023MWent}. They realized a photon-number entangled state between a microwave and optical modes based on the $\omega_p\rightarrow\Omega_p,\,\omega_p-\Omega_p$ SPDC in a \LN{} resonator. Here $\omega_p=193.46$ THz is the optical pump frequency and $\Omega_p=8.799$ GHz is the optical FSR, which is equal to the microwave frequency. The squeezing figure of -0.72 dB was observed. The experiment required a deeply cryogenic environment to prevent a thermal contamination of the microwave signal.  

\subsection{Quantum entanglement}\label{sec:entangle}

In the limit of very low optical power parametric light can be considered as a sparse flow of the signal-idler photon pairs. Then the two-mode continuos-variables entanglement reduces to the entanglement of a single photon pair. This pair can be entangled in various parameters, such as e.g. frequency, polarization, path, and even in more than one parameter at once (in which case one has \textit{hyper-entangled} states \cite{Suo:2015fv}).

Optical microresonators provide a promising platform for generating and manipulating quantum-entangled states of light. %Their small mode volumes and high optical quality factors enable strong field confinement and long photon lifetimes, enhancing nonlinear interactions while minimizing dissipative losses. These properties 
They have enabled outstanding results in the generation of %nonclassical optical states, including correlated and 
entangled photon pairs through spontaneous four-wave mixing and parametric down-conversion. The resonant enhancement of optical fields allows these processes to occur efficiently at relatively low pump powers, facilitating the development of compact quantum light sources. Combined with PICs, microresonators offer a pathway toward scalable quantum photonic technologies for entanglement-enabled communication and precision metrology. Their compatibility with atomic systems, solid-state quantum emitters, and microwave resonators creates further opportunities for hybrid quantum interfaces. Key challenges on this path include reducing optical losses and parasitic nonlinear interactions, improving photon-extraction and off-chip collection efficiencies, and achieving stable control of coupled resonator modes. %Addressing these issues will be essential for translating the demonstrated capabilities of microresonators into practical, scalable quantum photonic systems.

\paragraph{Polarization entanglement} of WGM photons is very desirable for quantum communication and information processing algorithms. Such algorithms have already been tested using polarization entanglement from less versatile photon-pair sources. On the other hand, tunable, narrow-band, single-mode WGM photons are well suited for coupling to optical transitions in atoms, ions, and other massive qubit implementations. This makes the polarization-entangled WGM-generated photon pairs a valuable resource for quantum information algorithms. Unfortunately, achieving such a polarization entanglement is difficult because of fixed polarization states of the signal and idler modes. It is virtually impossible to find a pair of e.g. TE polarized signal and idler modes at the same frequencies as a pair of TM polarized modes, and at the same time have those two pairs phase-matched to a single pump mode. 

One possible workaround is to use two nearly-identical but individually tunable resonators, one phase-matched for the TE signal and idler modes and the other for the TM modes, coupled to a common waveguide. This approach was realized by \cite{zhang2024PolEnt} with lithographically fabricated silicon nitride micro-rings. 

Instead of using two identical resonators, \cite{huang2024polEnt} relied on the directional degeneracy of WGMs. In this case the clockwise and counter-clockwise propagating signal and idler photons had the same polarization (TM), but rotating their polarizations and combining them on a polarizing beam splitter outside of the resonator allowed the authors to achieve a polarization-entangled state
\begin{equation}\label{polent}
|\Phi\rangle=(|HH\rangle+e^{i\phi}|VV\rangle)/\sqrt{2},
\end{equation}
where $H$ and $V$ indicate the horizontal and vertical polarizations, respectively. Their order corresponds to the detection channel. The phase $\phi$ depends on the relative path-lengths and can be used to vary the state $|\Phi\rangle$ between the Bell states $|\Phi^+\rangle$ and $|\Phi^-\rangle$. Notably, the signal and idler beams were nearly single-mode, witnessed by the normalized Glauber auto-correlation function $g_{ss}^{(2)}(0)\approx$ 1.9 (1.8) for the CW (CCW) signal photons. A more than six-sigma violation of the Clauser–Horne–Shimony–Holt’s inequality was achieved.

\paragraph{Frequency entanglement} of non-degenerate signal and idler modes was demonstrated by \cite{chen2025feq-ent}. Their setup also used counter-propagating WGMs combined outside of the resonator on a dichroic mirror to produce an entangled state 
\begin{equation}\label{wlent}
|\Phi\rangle=(|\lambda_1\lambda_2\rangle+e^{i\phi}|\lambda_2\lambda_1\rangle)/\sqrt{2},
\end{equation}
where $\lambda_1=909.6$ nm and $\lambda_2=1281.6$ nm are the signal and idler central wavelengths and their order again corresponds to the detection channel. While characterizing the polarization-entangled state (\ref{polent}) can be performed by rotating the wave plates and polarization analysers, characterizing the frequency-entangled state (\ref{wlent}) requires varying the optical path difference in a Hong-Ou-Mandel interferometer. In this way more that 87\% contrast two-photon interferometric fringe was observed, attesting to a high degree of the frequency entanglement.

Frequency entanglement of a photon pair can span more than two frequency values, in which case the state (\ref{wlent}) has more than two terms. Nearly equidistant WGM spectrum provides an optimal platform for such higher-dimensional entanglement. For example, \cite{Kues17ent} demonstrated the frequency entanglement among 10 signal and 10 idler eigenfrequencies leveraging the degenerate four-wave mixing (also known as the hyper-parametric conversion) in a micro-ring resonator fabricated on a chip from a high refractive index glass. Each entangled state term was individually manipulated, and a 10$\times$10 density matrix was tomographically reconstructed.
 
\paragraph{Multi-photon entanglement.} 
Not only photon pairs, but also triplets, quadruplets, and even larger numbers of photons can be entangled at the same time. They can be entangled in different ways, providing a variety of multi-photon entangled states important in fundamental quantum optics  as well as in quantum information processing \cite{Raussendorf01qcomp,Yeo2006,Agrawal06W}. 
% graph \cite{Hein04graph} states, Smolin states \cite{Smolin01}. 

Generating higher photon-number states requires higher-order optical nonlinearity, which is usually weak. This is where the high-$Q$ micro-resonators made from nonlinear optical materials can be particularly beneficial. A PIC comprising silicon microresonators and waveguides was theoretically studied by \cite{banic2024entangled} as a source of three-photon W-states \cite{Dur2000W,Eibl04W,Jin14W} and four-photon GHZ-states \cite{Greenberger90GHZ,Pan2000GHZ} based on hyper-parametric conversion. Substantial rates of such triplets and quadruplets reaching up to 10 kHz were predicted. 

Another approach to generation of entangled photon triplets is to harness a third-order nonlinear process which may be called 3PDC. It is analogous to the usual quadratic SPDC, but instead of a pair, a triplet of photons is borne out of an annihilated pump photon that share its energy (frequency) and momentum. The theory of resonator-assisted 3PDC is advanced by \cite{gonzalez2018_3PDC,denys2021_3PDC}. Distinctly from SPDC, \cite{gonzalez2018_3PDC} predict the absence of quadrature entanglement between the 3PDC photons in the spontaneous conversion regime. However the entanglement arises in the fully seeded regime.

Entangling or coupling of an even larger number of photons can lead to realization of advanced quantum computing and simulation algorithms, such as e.g. an all-optical Ising machine \cite{yamamoto2017Ising,tiunov2019Ising}. While a ``conventional" Ising machine is based on a network of coupled spin-1/2 particles, its photonic equivalent can be based on coupled OPOs, which also has two stable states in the phase space. This analogy was exploited by \cite{wang_coherent_2013} who parametrically coupled pulses circulating in a resonator, and later by \cite{Inagaki16Ising} who used a degenerate hyper-parametric conversion of dichromatic pulses. Each pulse was treated as a mode with its own ``spin value", i.e., the phase. Over 10 000  such modes were coupled, leading to photonic spin domains mimicking Ising spin arrays. The concept of an Ising machine based on an array of coupled micro-resonators was discussed and partially tested by \cite{okawachi2020Ising}. While multi-photon entanglement and resonator-based optical Ising machines offer promising routes toward more complex quantum systems, their further development will depend on improving nonlinear interaction strengths, generation efficiencies, and the scalability of coupled-resonator architectures.

\subsection{Photon pairs and heralded photons}\label{sec:herald}

Photon pairs generated in SPDC or spontaneous hyper-parametric conversion can also be used as a source of heralded single-photon states. In this approach a detection of e.g. an idler photon announces a presence of a signal photon. A statistical distribution of photon pairs (Poissonian in multi-mode case, thermal in single-mode case) permits for two or more signal photons per one heralding event, but in a low-flux regime such events can be made very rare. Then heralded photons can be used as an equivalent of a Fock state $|1\rangle$, which is important in quantum optics and communication algorithms. The probability of getting more then one photon per a heralding event can be characterized by the \textit{conditional} Glauber function $G^{(2)}_{ss|i}(0)$ inferred from a Hanbury Brown Twiss measurement in a signal channel conditioned on the idler photon detection. The lower this value, the better heralded photon approximates the Fock state. 

WGM and micro-ring resonators have made a good progress as photon-pairs and heralded-photon sources over past 10 years. SPDC photon pairs were generated in an x-cut lithium niobate micro-resonator with the photon-pair rate efficiency 5.13 MHz/$\mu$W \cite{xu2022pairs}. This enabled a very pure single-photon states with $G^{(2)}_{ss|i}(0)=0.01\pm 0.002$ with using the heralding. A 60 $\mu$m-diameter on-chip silicon carbide elliptical micro-ring was used for the same purpose by \cite{li2024SiC}, who reported the rate efficiency of 1.32 kHz/$\mu$W and an impressive $G^{(2)}_{ss|i}(0)=7\times 10^{-4}$. A rate efficiency of 5.8 kHz/$\mu$W was reported by \cite{guo2017AlN} in an on-chip aluminum-nitride micro-ring resonator.

Hyper-parametric conversion also was employed as photon-pair and heralded photon source. A 9 $\mu$m diameter silicon micro disc pumped with 73 $\mu$W at 1515.6 nm was used by \cite{lu2016heralded} to reach $G^{(2)}_{ss|i}(0)<0.05$. A 36 $\mu$m diameter silica microtoroidal resonator pumped with approximately 46 $\mu$W at 842.5 nm has yielded $G^{(2)}_{ss|i}(0)\approx 0.57$ \cite{hashimoto2020pairs}. Interference of heralded photons generated in two different micro-ring silicon resonators was demonstrated by \cite{faruque2018herald}.

Further intriguing possibilities open up with generation of heralded \textit{photon pairs}. This was demonstrated by \cite{huang2025HOM} who used counter-propagating WGM modes to obtain spectrally identical signal photons. The correlation function for two signal photons in a Hong-Ou-Mandel interferometer measurement conditioned on detecting two idler photons reached $G^{(2)}_{ss|ii}(0)\approx 0.2$.

The photon-pair rate is limited by the inverse of the optical coherence time. Upon reaching this limit the photon pairs lose their ``individuality" and we enter the squeezed light regime discussed earlier. This transition occurs near the OPO threshold and has a continuous character, accompanied by spectral and temporal-correlation changes of the emitted photon pairs. This behaviour was studied theoretically and experimentally by \cite{shafiee2020subthreshold}.

\subsection{Photon blockade}\label{sec:blockade}

Photon blockade is a nonlinear optical phenomenon when one optical field prevents another from coupling to a resonant system, such as an atomic transition or an optical resonator. At the quantum level, the photon blockade is equivalent to changing the photon statistics towards anti-bunching characterized by $G^{(2)}(0)<1$. 

Photon blockade is important particularly because it enables all-optical switches \cite{moroney2020gate}. An ultimate all-optical switch operated by a single photon can realize quantum logic gates, such as a controlled-not or a Fredkin gate. These are highly desirable but not yet achieved elements of an optical quantum computer. Their implementation with nonlinear WGM resonators requires entering the strong quantum regime when the non-linear conversion rate $g$ exceeds the linear loss rate $\gamma$. The state-of-the-art ratio of these rates experimentally observed by \cite{Lu2020PPLN-SH} in a small periodically poled lithium niobate resonator is $g/\gamma\approx 0.01$. The theoretical estimates may suggest higher values, but the experimental results usually fall quite short of that, e.g. $g/\gamma\approx 0.1$ (theory) vs. 0.002 (experiment) in case of \cite{Strekalov14switch}. 

The conventional photon blockade is based on the anharmonicity of the energy levels in a strongly nonlinear system. While the theoretical research in this direction continued for a variety of model systems \cite{wang2020blockade,zhou2020blockade,chen2022blockade}, the experimental progress has been limited by the requirement of a strong anharmonicity, which is difficult to fulfil in optics.

In conventional photonic blockade, the roles of the control and signal fields are typically symmetric, distinguished mainly by the field strength. This symmetry can be broken, and the blockade can be made non-reciprocal, by rotating the optical cavity as proposed by \cite{wang2019blockade,xue2020blockade,silver2021gyro,liu2023blockade,zheng2025blockade}. Such a non-reciprocal blockade can be viewed as a close analogy of the Fizeau drag. A distinct feature of the non-reciprocal photon blockade is that while the actual blockade occurs in the system pumped in one direction, the photon bunching arises for the opposite pumping direction. This approach can be based on the optical parametric amplification (OPA) using the quadratic nonlinearity \cite{wang2019blockade}, or coupling to a single-atom \cite{xue2020blockade,jing2021blockade,zheng2025blockade}, but can also be based on the cubic nonlinearity \cite{huang2018blockade,woodley2018Kerr} where it arises via the self- and cross-phase modulation \cite{hill2020symmetry}.

Meanwhile, a novel ``unconventional" photonic blockade concept has emerged  \cite{flayac2017unconventional,sarma2017blockade}, where the blockade arises from a destructive interference of the two-photon amplitudes corresponding to different conversion channels with both gain and dissipation present. The unconventional photonic blockade can be realized even with a weak single-photon nonlinearity. This approach was experimentally demonstrated by \cite{snijders2018blockade} in a micro-pillar cavity coupled to a single quantum dot. It was also theoretically studied in a bichromatically pumped single-mode resonator \cite{li2022blockade} and in a microring resonator with quadratic nonlinearity and counter-propagating coupled modes \cite{liu2023Parametric}. The quantum-optics aspect of the unconventional photon blockade was theoretically studied in a photonic molecules \cite{shen2019stat,shen2026BorB}. By controlling the parametric gain or dissipation in each resonator, a variation of the emitted light photon statistics from bunching  $G^{(2)}(0)>1$ to anti-bunching  $G^{(2)}(0)<1$ was predicted. The connection between non-reciprocal photon blockade and optical squeezing has been discussed by \cite{tang2022sqz-NPB,wang2023sqz-NPB}.

Finally, it is possible to realize a photonic blockade that is unconventional and non-reciprocal at the same time. Several architectures for its implementation have been theoretically proposed, breaking into two groups. In the first group, the non-reciprocity arises from the resonator rotation. The mode coupling in this case can be parametric \cite{shen2020blockade,gou2023blockade}, opto-mechanical \cite{li2019blockade}, or via coupling to atoms \cite{jing2021blockade,zhang2023blockade}. In the second group, the rotation is not required, and the non-reciprocity arise e.g. from interaction with a magnon \cite{xie2022magnon}. It can also be achieved via the destructive interference of two parametric amplifiers, as mentioned earlier. These could be implemented as the WGMs of two coupled resonators \cite{zhang2024blockade}, or as the CW and CCW WGMs of the same resonator \cite{liu2023Parametric,yuan2024blockade}. The arising of non-reciprocity is resonator without rotation or involving magnetic pseudo-vectors is a curious phenomenon, which may be traced to a spontaneous breaking of a parity-time symmetry \cite{del_bino2017PT,woodley2018Kerr}. It has been proposed \cite{del_bino2018PT} for the implementation of microresonator-based optical isolators and circulators.

\subsection{Coupling of WGM light to other quantum systems}

Coupling the resonator light to non-optical quantum system such as atoms, ions, NV centers, quantum dots, and others, potentially leads to quantum sensors and logic gates. In our previous review \cite{Strekalov16rev} we discussed the status and perspectives of this field, paying a special attention to coupling the WGM photons to atomic transitions. A more recent development in this field was reported by \cite{surya2018SHGtoRb} who excited a two-photon clock transition 5S$_{1/2}$--5D$_{5/2}$ in $^{85}$Rb with 778 nm light generated in an AlN microring resonator via the SHG process. The on-chip resonator had a 60 $\mu$m diameter and a loaded $Q$-factor of $3\times 10^5$.

An interesting research also was reported on magnons -- the quasi-particles describing the spin waves in ferromagnets or antiferromagnets. Magnons couple to light via the gyrotropic interaction, with the coupling strength proportional to the Verdet constant of the material. In optics, the yttrium iron garnet (YIG) is of a particular interest because it can simultaneously support relatively high-$Q$ optical WGM and magnons. Optically coupling a micro-sphere made from this material, \cite{Osada16magnon} measured $Q_{\rm mag}\approx 3000$ and the near-IR $Q_{\rm opt}\approx 1\times 10^5$. The sphere was magnetized perpendicularly to the WGM plane, which made the magnon frequency an approximately linear function of the bias magnetic field. A non-reciprocity was observed when the clockwise probe was back-scattering by magnons in the counterclockwise direction, but not vice versa. Note that the optical quality factor of YIG micro-spheres can be higher, reaching $6-8\times 10^5$ \cite{Zhang16magnon}. This work also reported a microwave-to-optic conversion mediated by magnons.

Interaction of the WGM light with magnons can be mediated by the spin-orbital coupling. In small resonators, the TM electric field is known \cite{Junge13} to have a significant longitudinal polarization $E_\phi$, coupling to the $\sigma_\pm$ transitions. The TE fields, on the other hand, are almost perfectly $\pi$-polarized with respect to the resonator axis $z$. Therefore the light scattering by a magnon leads to the TM-TE coupling, as was observed by \cite{Zhang16magnon}. In the strong opto-magnonic coupling regime this leads to their two-mode squeezing, theoretically described by \cite{xie2023ent_magnon}. 

Another important quasi-particle that can strongly couple with the WGM photons is a plasmon. Plasmons are collective excitations of free electrons. Surface plasmons usually have frequencies in the near-infrared to visible range, and can directly interact with the resonator's evanescent fields. They furthermore can have a sufficiently low loss to form their own high-$Q$ WGMs. However the research of the optical-plasmon WGM coupling remains largely theoretical \cite{heylman2016plasmon,kang2017plasmon,tiwari2021plasmon}.

\subsection{Quantum Nondemolition Measurements} \label{sec:QND}

Quantum nondemolition (QND) measurements represent a unique class of quantum measurements in which the quantum observable of interest can be measured repeatedly without perturbing its subsequent evolution \cite{Braginsky1980Science}. In a QND measurement, the inevitable measurement back-action required by quantum mechanics is diverted into a conjugate variable, leaving the measured quantity unchanged. Consequently, repeated measurements of the same observable yield identical results, limited only by the intrinsic dynamics of the system rather than by the measurement process itself. 

One of the most widely discussed optical implementations of a QND measurement is based on the optical Kerr effect. Consider a signal beam containing an unknown number of photons propagating through a nonlinear medium together with a probe beam. The third-order optical nonlinearity causes cross-phase modulation between the two fields, such that the refractive index experienced by the probe depends on the intensity of the signal field. As a result, the probe acquires a phase shift proportional to the signal photon number. By measuring this phase shift it becomes possible to infer the photon number without absorbing, scattering, or otherwise destroying the signal photons. In the ideal limit, the signal photon number is preserved, enabling subsequent measurements and satisfying the defining criterion of a QND measurement.

The practical realization of such measurements is challenging because optical nonlinearities at the single-photon level are inherently weak. In most conventional optical media, the phase shift induced by a single photon is many orders of magnitude smaller than both technical and quantum phase noise sources. Optical losses, including absorption, scattering, and imperfect coupling, directly destroy photons and therefore violate the nondemolition criterion. Fluctuations associated with laser frequency noise, thermal effects, and vacuum fluctuations often overwhelm the nonlinear signal. The challenge is to engineer systems in which the nonlinear interaction strength is enhanced while maintaining extremely low dissipation and low excess noise. 

The combination of long photon lifetime and strong spatial confinement in WGM resonators significantly enhances Kerr nonlinear processes, making WGM resonators one of the most promising platforms for quantum nonlinear optics and QND measurements. The theoretical foundations of optical QND measurements in resonators have been discussed extensively in the literature and summarized in earlier reviews \cite{Grangier1998QND}. More recently, Ref.~\cite{balybin2022QND}  extended this framework by performing a detailed analysis of the noise processes that ultimately limit measurement sensitivity and by proposing realistic experimental implementations based on WGM resonators. The study identified the conditions under which the nonlinear phase shift generated by a single intracavity photon can exceed the combined contribution of shot noise and technical noise. It also highlighted the critical role of intrinsic cavity loss, coupling loss, and thermodynamic fluctuations in determining the feasibility of QND detection. It was shown more recently that the intracavity squeezing of the probe beam in the QND scheme \cite{balybin2022QND} enables cancellation of the SPM effect \cite{Salykina2025Intracavity}.  A squeezed quantum state of the probe beam and the antisqueezing of this beam at the output of the microresonator also can improve the measurement sensitivity \cite{PhysRevA.108.053708}. 

Among the fundamental noise sources, thermorefractive noise occupies a particularly important position. Random temperature fluctuations within the resonator material induce fluctuations of the refractive index, which in turn generate resonance-frequency noise. Thermorefractive noise has been identified as one of the principal fundamental limitations affecting frequency stability, sensing performance, and quantum measurements in WGM systems. Thermal effects can also lead to complex optothermal dynamics, resonance drifts, and nonlinear instabilities that must be carefully controlled in precision quantum-optical experiments.

The broader significance of QND measurements extends well beyond photon counting. QND techniques provide a foundation for quantum feedback control, quantum error correction, entanglement generation, and quantum-enhanced sensing. They have played a central role in cavity quantum electrodynamics experiments, including the nondestructive detection of single photons in microwave cavities and the repeated observation of photon-number quantum jumps. More generally, the pursuit of optical QND measurements continues to drive the development of ultra-low-loss resonators, quantum-limited photonic devices, and novel nonlinear materials. In this context, high-Q WGM resonators remain among the most promising architectures for realizing practical nondestructive measurements of light.

\subsection{Nongaussian states}

Nonlinear optical microcavities provide a powerful platform for the generation and manipulation of nonclassical states of light, owing to their ability to simultaneously enhance optical nonlinearities and reduce dissipative losses \cite{Clemmen2009,Azzini2012,Engin2013,Guo2014,Grassani2015,Wakabayashi2015,Suo:2015fv,zhao2020sqz}. These ideas resulted in propositions of a broad range of quantum photonic functionalities, including photon-pair generation, production of entangled states, quadrature squeezing, and engineering of highly non-Gaussian optical fields. The combination of ultra-high quality factors and small mode volumes available in modern microresonators significantly strengthens light-matter interactions, making such systems attractive candidates for deterministic quantum state preparation.

One particularly promising approach relies on the Kerr nonlinearity of an optical cavity. It has been shown that a Fock state can be deterministically generated in a Kerr resonator through an adiabatic evolution of the system parameters, specifically the cavity detuning and the amplitude of the external driving field \cite{PhysRevA.100.033822}. During the adiabatic process, the quantum state follows the instantaneous eigenstate of the driven nonlinear system, ultimately evolving into a state with a well-defined photon number. Because the protocol does not rely on precise timing or abrupt switching events, it exhibits intrinsic robustness against fluctuations of the driving field and imperfections in the control sequence. This feature makes the method particularly attractive for practical implementations in realistic experimental environments.

The flexibility of nonlinear photonic systems extends beyond the generation of single Fock states. More sophisticated quantum state engineering can be realized in coupled nonlinear resonator architectures, where multiple interacting cavities provide additional degrees of freedom for tailoring quantum dynamics \cite{Rivera2023FockBIC}. An especially intriguing concept introduced in this context is the notion of an $n$-photon bound state in the continuum. In such a system, a collective resonant optical mode is engineered so that radiative losses vanish for a specific intracavity photon number $n$, while states with different photon populations remain lossy. This highly nonlinear mechanism enables selective stabilization of particular quantum states and opens new avenues for the generation of large-photon-number Fock states and strongly number-squeezed optical fields. The approach illustrates how carefully engineered dissipation can become a resource for quantum state preparation rather than a limitation.

Despite the progress in generating nonclassical intracavity states, an equally important challenge is the faithful transfer of these states from the resonator into propagating optical modes. The process of out-coupling light from a cavity is not necessarily benign. It can alter the quantum properties of the emitted field and degrade the nonclassical features that were established inside the resonator. Theoretical studies have demonstrated that the coupling conditions, external losses, and modal overlap between cavity and output channels critically influence the fidelity of the extracted quantum state. This issue is particularly relevant for Fock states, Schr\"odinger-cat states, squeezed states, and other non-Gaussian states whose quantum signatures are highly susceptible to loss and mode mismatch. Consequently, practical realization of microcavity-based quantum light sources requires simultaneous optimization of both intracavity state generation and cavity-to-waveguide state transfer.

An experimental framework for observing and characterizing nonclassical states emitted from high-Q Kerr microresonators was proposed in \cite{Balybin2020QuasiFock}. The study analyzed the influence of intrinsic cavity losses, output coupling losses, detector imperfections, and technical noise originating from classical optical sources used in the measurement process. The results indicate that, despite these practical limitations, the generation and observation of strongly non-Gaussian optical states outside the resonator remain achievable with realistic device parameters. In particular, the analysis suggests that modern ultra-high-Q microresonators can support the production of quasi-Fock states exhibiting pronounced nonclassical characteristics, including photon-number squeezing and non-Gaussian phase-space distributions. 

\section{Summary and conclusion}\label{sec:Summary}

We reviewed the progress of whispering gallery and similar types of optical resonators over the past decade, focusing on their applications in nonlinear and quantum optics. As anticipated in our 2016 review \cite{Strekalov16rev}, these resonators have evolved from a laboratory curiosity to building blocks of modern optical technology. One testimony to this is the number of commercial companies embracing the microresonator technology. In addition to the earlier established players such as Menlo Systems, TOPTICA Photonics, OEwaves, and Pilot Photonics, we now have HyperLight, Anello Photonics, LioniX International, Alpes Lasers, Solinide Photonics AB, Octave Photonics, Enlightra Quintessent, LIGENTEC, and likely more by the time this review reaches the readers.

In our 2016 review we outlined the major obstacles to the progress in the field, such as weak optical nonlinearity, strong thermal and mechanical sensitivity of the stand-alone resonator systems, and their poor scalability. For the emerging on-chip technology, the inadequate surface quality was preventing the high quality factors. Several possible thrust directions were then identified that may solve or circumvent these problems. It is interesting to see in the retrospect how this played out, and project the further development in the field.

The progress of the quality factor of the on-chip TFLN resonators is captured in Table~\ref{tab:LNprogress}. While most of its progress is associate with the surface quality improvement, the 2025 result by \cite{barya2025slow_light} is different. Here, the spectral hole burning facilitated a strong dispersion dramatically increasing the optical length of the resonator within a narrow frequency window, without changing its physical size. Therefore the loss rate per unit \textit{optical} length is greatly decreased, and the $Q$-factor increased. These improvements may be coming at a slow pace, but there are no indications of any fundamental hurdle preventing the achievement of the ``perfect" surface quality when the Rayleigh loss effect on the $Q$-factor no longer matters. Hand in hand with improving the quality factors of the PIC micro-resonators goes the size, quality and versatility of the resonator arrays, such as discussed in Section~\ref{sec:oalas}.

%%%%%%%%%%%%%%%%%%%%%%%%%%
\begin{table*}[htb]
\centering
\caption{The progress in thin-film lithium niobate on-chip resonators intrinsic quality factor $Q_0$}.
\label{tab:LNprogress}
\begin{tabular}{p{1.4cm}p{1.7cm}p{1.4cm}p{3.1cm}p{2.2cm}p{4.8cm}}
\hline\\
\textbf{Year} & 
\textbf{Best $Q_0$} &
\textbf{Ref.} & 
\textbf{Resonator type} &
\textbf{Coupler} &
\textbf{Fabrication method} 
 \\
\hline\\

2016 &
$2.7\times 10^5$ & %  intrinsic
\cite{jiang2016optomech} &
Z-cut microdisk &
Tapered fiber &
E-beam lithography, reactive ion dry etching  \\

2017 &
$1\times 10^7$ &
\cite{zhang_monolithic_2017} &  %  intrinsic; loaded = 5e6
X-cut microrings, racetracks &
Integrated waveguide &
E-beam lithography, reactive ion dry etching  \\

2018 &
$>1\times 10^7$ &    %  intrinsic; loaded = 3e6
\cite{wuLithiumNiobateMicrodisk2018} &
X-cut microdisk &
Tapered fiber &
Fs-laser patterning, chemo-mechanical polishing \\

2021 &
$>1\times 10^8$ &  %  intrinsic; loaded = 8e7
\cite{gaoLithiumNiobateMicroring2021} &
X-cut microring &
Tapered fiber &
Chemo-mechanical etching and polishing \\

2023 &
$4\times 10^7$ &   %  intrinsic; loaded = 4e6
\cite{li2023highQ} &
Z-cut microring &
Integrated waveguide &
Photolithography-assisted chemo-mechanical etching, annealing  \\

2024 &
$2.9\times 10^7$ &     %  intrinsic
\cite{zhu2024highQ} &
Racetrack &
Integrated waveguide &
E-beam lithography, reactive ion dry etching, annealing  \\

2025 &
$1.1\times 10^8$ &
\cite{barya2025slow_light} &
Er-doped microring &
Integrated waveguide &
E-beam lithography, reactive ion dry etching, annealing, spectral hole burning  \\

\hline
\end{tabular}
\end{table*}
%%%%%%%%%%%%%%%%%%%%%%%%%

The research of novel low-loss, high-nonlinearity resonator materials also shown a significant progress, as discussed in Sections~\ref{sec:fab} and \ref{sec:chip}. One highlight is the silicon nitride discussed in Section~\ref{sec:Si3N4}, which has been declared a new versatile manufacturable platform for quantum photonics \cite{psiquantum_team2025Nat}. Another good candidate is barium titanate \ce{BaTiO3}, a perovskite with a strong electro-optical response \cite{psiquantum_team2025Nat}. Besides the traditional emphasis on the visible light, near- and mid-IR materials, the UV and deep UV resonator-compatible crystals are now a large and active research direction \cite{kang2022DUV,li2023nitrde,zhang2024re,yan2024crystals}.

The rapid progress in fabrication, dispersion engineering -- now with the help from AI \cite{pal2023AIdesign}, -- and heterogeneous integration suggests that integrated WGM and ring resonators are evolving from specialized laboratory components into a general-purpose technology for nonlinear photonics, microwave photonics, metrology, and quantum systems. Rather than converging on a single “best” material, the field is differentiating into complementary platforms. For instance, silica can be used for ultimate low loss and ultra-high-Q, \ce{Si3N4} for scalable Kerr photonics, LN and LT for electro-optic microwave-optical interaction, and AlN for piezoelectric and visible-to-UV applications. The emerging platforms with strong optical nonlinearities, such as SiC, GaP, AlGaAs, diamond, and GaN, can enable quantum functionality and provide the access to new wavelength ranges. These wide material choices is a strength of the field, which enables it to be tailored to specific scientific and technological applications.

This progress was expected to lead to efficient all-optical switches, ultimately achieving the strong interaction between individual photons (also referred to as the single-photon anharmonicity), $g>\gamma$. This would enable a realization of the quantum logic gates, which would be a revolutionary step in quantum-optical computing. The progress towards this goal is reviewed by \cite{Lu2020PPLN-SH}, but as we mentioned in Section~\ref{sec:blockade}, achieving this ambitious goal so far remains elusive. While the best realistic theoretical analysis projects $g/\gamma\approx 0.1$, which is still an order of magnitude short of the benchmark, the experimental results are typically at least another order of magnitude lower. This discrepancy is almost universal in the WGM-based nonlinear optics, and solving it may be one of the future research and technology goals in the field.

An alternative to gate-based optical quantum computing is the analogue quantum computing and simulation. One approach in this field is using the OPO-based optical Ising machine discussed in Section~\ref{sec:entangle}. This approach shows some experimental progress \cite{Inagaki16Ising,okawachi2020Ising} in terms of multipartite optical modes entanglement and holds a promise for the future technology. It is likely to be pursued further.

Finally, the optical and microwave metrology as well as microwave photonics will likely remain one of the driving forces in the area. One important direction is the realization of next-generation low-noise microwave sources based on optical frequency division and nonlinear frequency conversion. Ultra-high-Q WGM resonators can support stable optical modes with extremely narrow bandwidth, enabling photonic oscillators with significantly reduced phase noise and improved long-term stability compared with the best conventional electronic oscillators. Integration of WGM resonators with semiconductor lasers, microcombs, and microwave extraction circuits will lead to compact photonic microwave synthesizers capable of approaching the performance of laboratory-scale references while operating with substantially reduced size, weight, and power consumption. There has been a significant progress in the phase-locked broad optical Kerr combs underlying this field, reviewed in Section~\ref{sec:comb}. Particularly promising may be the $\chi^{(2)}$ combs discussed in Section~\ref{sec:PDC}, or their combinations with the Kerr ($\chi^{(3)}$) combs. 

Another promising development is the use of WGM resonators as interfaces between optical and microwave domains. The strong optomechanical, electro-optical, and nonlinear interactions available in carefully engineered WGM structures can enable efficient frequency conversion, coherent signal transfer, and quantum transduction between microwave and optical photons. Such capabilities are important for future communication systems, distributed sensing, and quantum networks. In addition, tunable WGM resonators with engineered dispersion and coupling properties will provide reconfigurable microwave photonic filters, delay elements, and signal processors with unprecedented selectivity and stability, enabling adaptive RF systems for applications ranging from deep-space communication to advanced radar and navigation.

A further important development is expected from combining WGM resonators with integrated photonics platforms. Hybrid architectures combining crystalline WGM resonators, low-loss waveguides, and active materials could merge the exceptional performance of bulk resonators with the scalability of chip-based systems. This may result in compact frequency references, microwave photonic processors, and precision sensors capable of operating in harsh environments. Beyond classical applications, WGM-based microwave photonic systems may also become key components in quantum technologies by enabling low-noise interfaces, generation of nonclassical microwave states, and controlled interactions between optical, microwave, and mechanical degrees of freedom.

%\newpage

\ack
Research conducted by ABM was carried out at the Jet Propulsion Laboratory,
California Institute of Technology, under a contract
with the National Aeronautics and Space Administration
(80NM0018D0004).
Research conducted by FS was funded by Horizon Europe, Marie Sklodowska-Curie Actions (MSCA) (Grant No. 101154422).

%\newpage
%\section*{References}

%\bibliographystyle{alpha} %use this for alphabetic sorting and removing of double references
\bibliographystyle{iopart-num}
\bibliography{bibliography1, referencesFreeStanding, referencesIntegrated, bibliographyAM}
%\bibliography{referencesIntegrated}

\end{document}

%% file: transductionProgressFig.tex
%ref, year, internal efficiency,uncertainty
%
%Optomechanical
%andrews_bidirectional_2014, 0.086, 0.007
%bagci_optical_2014, 0.8, 0
%Higginbotham2018, 0.47, 0.01
%
%Electrooptic
%Fan:18, 0.259, 0.003
%Rueda:16, 0.0109, 0.0002
%Khan:07,3.7e10?9,0
%strekalov_efficient_2009, 2.6e10?6, 0

\begin{figure*}[h]
\centering

\begin{tikzpicture}
\begin{semilogyaxis}[line width = 1.2,
    grid=both,
		grid style={line width=0.8pt, draw=gray!30},major grid style={line width=0.4pt,draw=gray!50},
		width=15.5cm,height=8cm,
    xlabel={Year},
    xmin=2006,xmax=2026,
		ytick={1,1e-2,1e-4,1e-6,1e-8,1e-10,1e-12},
		yticklabels={$1$,$10^{-2}$,$10^{-4}$,$10^{-6}$,$10^{-8}$,$10^{-10}$,$10^{-12}$},
    ylabel={Efficiency},
    ymin=1e-14,ymax=1.4,
    legend pos=south west,
        legend cell align={left},
        legend style={draw=none, column sep=0.3cm},
		/pgf/number format/.cd,
        use comma,
        1000 sep={}
]

\addplot[
		only marks,
    color=red,
    mark=*,
		nodes near coords={\cite{\thecitation}},
    nodes near coords align={\alignment},
    visualization depends on={value \thisrow{cite} \as \thecitation},
		visualization depends on={value \thisrow{align} \as \alignment},
    every node near coord/.style={black}
    ]
		table[x=x, y=y]{eo.dat};
		
    \addlegendentry{Electro-optic}

\addplot[
		only marks,
    color=black,
    mark=o,
		nodes near coords={\cite{\thecitation}},
    nodes near coords align={\alignment},
    visualization depends on={value \thisrow{cite} \as \thecitation},
		visualization depends on={value \thisrow{align} \as \alignment},
    every node near coord/.style={black}
    ]
		table[x=x, y=y]{mo.dat};
		
    \addlegendentry{Magneto-optic}	
		
\addplot[
		only marks,
    color=green,
    mark=*,
		nodes near coords={\cite{\thecitation}},
    nodes near coords align={\alignment},
    visualization depends on={value \thisrow{cite} \as \thecitation},
		visualization depends on={value \thisrow{align} \as \alignment},
    every node near coord/.style={black}
    ]
		table[x=x, y=y]{lambda.dat};
		
    \addlegendentry{$\Lambda$- and Rydberg systems}
		
\addplot[
		only marks,
    color=blue,
    mark=*,
		nodes near coords={\cite{\thecitation}},
    nodes near coords align={\alignment},
    visualization depends on={value \thisrow{cite} \as \thecitation},
		visualization depends on={value \thisrow{align} \as \alignment},
    every node near coord/.style={black}
    ]
		table[x=x, y=y]{om.dat};
		
    \addlegendentry{Optomechanical}			

\addplot[
		only marks,
    color=gray,
    mark=*,
		nodes near coords={\cite{\thecitation}},
    nodes near coords align={\alignment},
    visualization depends on={value \thisrow{cite} \as \thecitation},
		visualization depends on={value \thisrow{align} \as \alignment},
    every node near coord/.style={black}
    ]
		table[x=x, y=y]{piezo.dat};
		
    \addlegendentry{Piezoelectric}

\addplot[color=red!60, dashed, line width=1.2 ] table[row sep = crcr]{2006 3e-7 \\ 2026 3e-7 \\};

\end{semilogyaxis}
\end{tikzpicture}

\caption{\textbf{Progress in microwave up-conversion efficiency.} Electro-optic techniques (red) were initially applied to non-resonant systems~\protect{\cite{Khan:07}} before resonant enhancement of, first, the optical field~\protect{\cite{Savchenkov09ol, Strekalov09THzOL}}, and then also the microwave field~\protect{\cite{rueda_efficient_2016, fan2018upconversion,warner_coherent_2025}}. Recent magneto-optical upconversion experiments (black circle) using WGMs in YIG~\protect{\cite{hisatomi, Osada16magnon, Zhang16magnon}}, while showing promise, still suffer from low efficiencies, although in future this may increase with the use of resonant microwave fields. First demonstrations based on rare earth ions~\protect{\cite{fernandez-gonzalvo_frequency_2015}} (green) were also low in efficiency, but this figure was later enhanced by using an optical cavity~\protect{\cite{PhysRevA.100.033807}} and on chip scale systems~\cite{xie_scalable_2025}. Recently, higher efficiencies have been achieved using Rydberg states in rubidium atoms~\protect{\cite{PhysRevLett.120.093201, PhysRevA.99.023832,tuHighefficiencyCoherentMicrowavetooptics2022}}. Optomechanical systems (blue) have shown the highest demonstrated efficiencies so far~\protect{\cite{andrews_bidirectional_2014, bagci_optical_2014, Vainsencher:16,jiangEfficientBidirectionalPiezooptomechanical2020}}, but often require laser cooling or feed-forward techniques to suppress thermal noise~\protect{\cite{higginbotham2018conv}}. Recent demonstrations have enhanced optomechanical interactions by using piezoelectric materials~\protect{\cite{hanCavityPiezomechanicsSuperconductingnanophotonic2020, stockillUltralownoiseMicrowaveOptics2022, weaverIntegratedMicrowavetoopticsInterface2023,meesala_quantum_2024}}. We also show the efficiency achievable by using a commercial electro-optic modulator (red dashed line).}
\label{fig:efficiency_progress}
\end{figure*}